# Advances in Optical and Microwave Nonreciprocity

Sergey V Kutsaev[1#], Alex Krasnok[2#], Sergey N. Romanenko[3], Alexander Yu. Smirnov[1], Kirill Taletski[1], and Vyacheslav Yakovlev[4]

[1]*RadiaBeam Technologies, LLC, Santa Monica, CA 90404, USA*

[2]*Photonics Initiative, Advanced Science Research Center, City University of New York, NY 10031, USA*

[3]*Zaporizhzhya National Technical University, Ukraine*

[4]*Fermi National Accelerator Laboratory, Batavia, IL 60510, USA*

#*These authors contributed equally to this work.*

## Abstract

Modern photonic and quantum technologies demand reciprocity breaking, e.g., isolators, full-duplex systems, noise isolation in quantum computers, motivating searching for practical approaches beyond magnet-based devices. This work overviews the up-and-coming advances in optical nonreciprocity, including new materials (Weyl semimetals, topological insulators, metasurfaces), active structures, time-modulation, PT-symmetry breaking, nonlinearity, quantum nonlinearity, unidirectional gain and loss, chiral quantum states, and valley polarization.

## Introduction

Modern classical and quantum technology rely on our ability to control electromagnetic waves' propagation and their classical and quantum interaction with matter. Over the last few decades of this journey, researchers have achieved genuinely staggering success. Examples include the invention of photonic crystals enabling tailoring light band-gap structure at our will[1,2], plasmonics with unprecedented capabilities for light localization in nano dimensions[3], all-dielectric nanophotonics[4,5], metasurfaces for low-loss and highly efficient lensing and sensing[6–8], and topologically nontrivial phases of light-matter interaction for robust photonic devices and scattering-less unidirectional wave propagation[9]. Underpinning these devices lies the idea of a "metamaterial" that is, in its broadest sense, sculpturing a material with certain properties on demand from "meta-atoms" and their special arrangement in a similar fashion as Nature creates an ordinary material but rather in a blind way.

Realizing that we are no longer limited to a somewhat narrow set of natural materials, but can design them at our will, made us rethink many fundamental concepts and even question facts that seemed impossible. Quantum computing is another example of how new ideas and concepts revolutionize the way we approach scientific problems[10]. To illustrate the difference between classical and quantum approaches, we can say that it is like solving a question of whether a complex maze has a way out. The classical approach relies on a blind search among all possible paths, which can take an excruciatingly long time. Instead, one can launch, say, an acoustic wave into the maze and, utilizing interference (analog of quantum coherence), test all possible paths at once. Nowadays, we have achieved stunning results, but in the future, such quantum computers will allow us to tailor materials at will using the concept of metamaterials instantly, avoiding blind searches.

Effective nonreciprocal materials are another product of the metamaterials concept. The general observation can express reciprocity in electromagnetics: "If I can see you, you can see me." This phenomenon gets its origin from the fact that physical laws governing electromagnetic fields, light in this example, and their interaction with the matter on a microscopic level do not prefer any particular direction of time in the absence of a magnetic field. In other words, if we were to reverse the direction of time hypothetically, the predictions of the physics laws would remain unchanged. Of course, this is not the case at a macroscopic scale because the arrow of time has a fixed direction, as we know from our daily experience. In practice, this means that the transmission coefficient between any two points in the system will be the same regardless of transmission direction.

In many practical applications, it is required that the waves, propagating through a media or a device, can exhibit different responses when source and observation points are interchanged. The examples include isolators preventing wave scattering back to lasers and generators. Another example is full-duplex systems for multiplexing transmission and receiving in the same channel without mutual interference. Preventing superconductor quantum computers working at very low temperatures from thermal noise that otherwise comes to the system from our equipment staying at the room temperature also relies on the nonreciprocal response[11].

However, today's nonreciprocal components are almost exclusively realized through the magneto-optic Faraday effect. This effect requires ferrite materials [ferrites, such as Yttrium Iron Garnet (YIG) and materials composed of iron oxides and other elements (Al, Co, Mn, Ni)] with static magnetic field bias. These devices are expensive, hard to tune, bulky and incompatible with planar technologies like silicon-based integrated circuits[12] and transmission-line quantum circuits[13,14] that power the wireless and computing revolutions. Also, ferrite materials suffer from strong dissipative losses interfering with their use, especially in optics. We would also add that the magnetic field breaks RF superconductivity, especially when we talk about single-photon levels. In fact, the fields <1 nT are required for superconducting qubits and SQUIDs, so the presence of a ferrite circulator implies the requirement of an additional magnetic shielding[15].

Recently novel avenues to achieve nonreciprocity of electromagnetic fields without magnets but using new scattering effects and a new class of materials and metamaterials have been implemented[16–21]. The examples include dynamic spatiotemporal modulation of parameters[22–25], synthetic magnetic field[25–27], angular momentum biasing in photonic or acoustic systems[21,28,29], nonlinearity[30–33], interband photonic transitions[34,35], optomechanics[36–40], optoacoustics[41,42], PT-symmetry breaking[43–45], unidirectional gain and loss[46–53], moving/rotating cavities[54–56] and emitters[57], Doppler-shift[58], chiral light-matter coupling and valley polarization[59–63], and quantum nonlinearity[64–67]. Furthermore, quantum systems based on superconducting Josephson junctions attract much attention as they hold a great promise for quantum computing[13,65,68–71]. Nevertheless, magnetic nonreciprocal systems continue to develop in relation to new materials (Weyl semimetals, topological insulators, metasurfaces) and effects[72]. Nonreciprocity of waves of other Nature is also of enormous interest today[73–75].

This work is dedicated to a review of the state-of-the-art advances in nonreciprocity. First, we briefly discuss the physical picture that lies behind nonreciprocal materials and systems. We provide the major characteristics that describe nonreciprocal systems. We briefly discuss the magnet-based nonreciprocal systems (NSs) and their disadvantages. Next, we focus on new materials for nonreciprocity, including Weyl semimetals, topological insulators, and metasurfaces. Then we survey NSs based on active structures, PT-symmetry breaking, time-modulation, nonlinearity, and quantum effects. Throughout the work, we discuss the pros and cons of the mentioned approaches to nonreciprocity.

## 1. Reciprocal and nonreciprocal systems

The reciprocity principle is one of the most fundamental concepts in Nature and stems from the fundamental time-reversal symmetry of physical laws. As we know, all microscopic physical laws obey time-reversal symmetry if all odd quantities, for instance, currents and magnetic field in the Maxwell equations, reverse their sign. The important case of dissipative macroscopic systems is discussed below.

Reciprocity implies that the wave transmission between any two points in an arbitrary system is the same for opposite propagation direction as far as the *time-reversal symmetry* preserved[19,21,76–82]. For example, let us consider an arbitrary complex time-reversal symmetric system and two arbitrary areas A and B. If there is a current $\mathbf{J}_A$ ($\mathbf{J}_B$) in area A (B), then this current creates some field in area B (A), Figure 1(a). The reciprocity dictates that the mutually induced powers of these two currents are equal:

$$\int_V \mathbf{J}_A \mathbf{E}_B dV = \int_V \mathbf{J}_B \mathbf{E}_A dV, \tag{1}$$

where integration runs over the area of non-zero current. In the case of point-like areas, this formula simplifies to $\mathbf{J}_A \mathbf{E}_B = \mathbf{J}_B \mathbf{E}_A$. In the case of linear RLC circuits, this condition can also be written as $I_2 V_1 = I_1 V_2$, where a voltage $V_1$ ($V_2$) being applied to the first branch of a circuit creates a current $I_2$ ($I_1$) in the second branch.

The reciprocity principle allows making predictions regarding any arbitrary complex systems' response and operation, simplifying the analysis. However, there are several practical situations in which it is advantageous to break reciprocity. Examples of such situations include (i) preventing back-scattering of signals back to sources (lasers, generators), (ii) full-duplex systems, (iii) nonreciprocal cavity (resonator, qubit) excitations and readout, (iv) breaking the time-reversed symmetry in emission and absorption (Kirchhoff's law of thermal radiation)[78,83], etc. For example, full-duplex systems are very beneficial for communications because they allow simultaneous transmission and reception of signals at the same frequency through the same antenna[84]. The nonreciprocal resonator excitation is promising for optical memory elements and energy harvesting. Nonreciprocal qubit readout would allow for noiseless control of a qubit state.

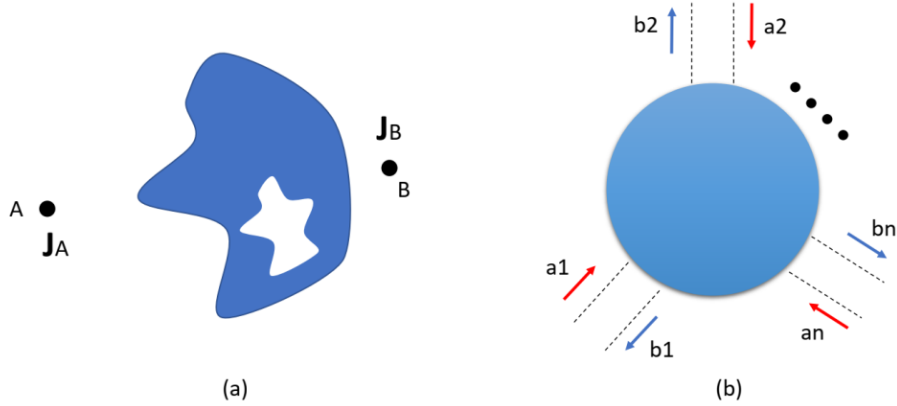

**Figure 1**. **Reciprocal and nonreciprocal systems.** (a) Arbitrary structure and two current sources A and B. The reciprocity principle dictates that the transmission A -> B and B -> A are equal. An alternative formulation is that the induced powers are equal, Eq.(1). (b) Equivalent n-port representation for the illustration of the S-matrix framework. The blue sphere denotes a general structure (cavity, circulator, etc.).

The time-reversal operation, $\hat{T}$, changes the arrow of time and hence reverses all processes that happen in a system. If a (scalar or vector) function $\mathbf{\Psi}(\mathbf{r},t)$ describes the electromagnetic field in a system, then the essence of the operator $\hat{T}$ is:

$$\hat{T} \cdot \mathbf{\Psi}(\mathbf{r},t) = \mathbf{\Psi}'(\mathbf{r},-t) . \tag{2}$$

If $\mathbf{\Psi}'(\mathbf{r},-t) = \mathbf{\Psi}(\mathbf{r},t)$, then, the system is time-reversal symmetric; otherwise, it is asymmetric. Since the direct and reverse parts of any process (wave propagation) describe the system's response for opposite transmission directions, time-reversal symmetry/asymmetry is inherently related to reciprocity/nonreciprocity[76,82]. All quantities we consider in electrodynamics are either symmetric (even), $\mathbf{\Psi}'(\mathbf{r},-t) = +\mathbf{\Psi}(\mathbf{r},t)$, or antisymmetric (odd), $\mathbf{\Psi}'(\mathbf{r},-t) = -\mathbf{\Psi}(\mathbf{r},t)$. The examples of even quantities are charge, E-field, polarization. The odd quantities are H-field, velocity and current, wavevector, and Pointing vector. Note that in the reciprocal frequency space, the time-reversal operation consists in taking *complex conjugation* and changing the sign of the odd quantities, for example, $\mathbf{E}_\omega(\mathbf{r}) \xrightarrow{\hat{T}} \mathbf{E}_\omega^*(\mathbf{r})$, $\mathbf{H}_\omega(\mathbf{r}) \xrightarrow{\hat{T}} -\mathbf{H}_\omega^*(\mathbf{r})$.

Eq. (1) gives us a clue that one should measure transmission coefficients to find out whether a system is reciprocal or not. The scattering matrix (S-matrix) concept provides a compelling approach to this end[85]. Let us consider a general system depicted in Figure 1(b) and define the condition under which the system is reciprocal. For this, we first chose a basis of certain incoming and outgoing waves, or *channels* with corresponding amplitudes $\mathbf{a} = \{a_1, a_2, ..., a_n\}$ and $\mathbf{b} = \{b_1, b_2, ..., b_n\}$, respectively. Channels exist outside the system and represent freely propagating solutions of the wave equation in the absence of the scattering object[85]. Hence, the formal solution of the wave equation in the basis of chosen channels can be presented in the form

$$\mathbf{b} = \hat{S}\mathbf{a}, \qquad (3)$$

where $\hat{S}$ is the scattering matrix of the system on the chosen basis. Eq.(3) simply expresses that in a linear system, amplitudes of the outgoing waves $\mathbf{b}$ are linear combinations of the incoming wave amplitudes $\mathbf{a}$ and connected by the $\hat{S}$ matrix of the system. For example, in a single port system, the $\hat{S}$ matrix coincides with the reflection coefficient ($r$). In a two-port system, the S-matrix is $\hat{S} = \begin{pmatrix} r_{11} & t_{12} \\ t_{21} & r_{22} \end{pmatrix}$, where $r_{ii}$ and $t_{ij}$ stand for corresponding reflection and transmission coefficients. This approach can be generalized to the multi-mode ports[82].

In a reciprocal system, according to Eq.(1), $t_{ij} = t_{ji}$, and hence the scattering matrix is symmetric,

$$\hat{S} = \hat{S}^T, \qquad (4)$$

Therefore, in order to test a system, whether it is reciprocal or not, one should measure or calculate the mutual transmission between each pair of channels. Breaking of time-reversal symmetry makes the S-matrix asymmetric, $\hat{S} \neq \hat{S}^T$.

Note that in bulky media, the nonreciprocity requires the electric permittivity ($\hat{\varepsilon}$) or magnetic permeability ($\hat{\mu}$) to be asymmetric, that is $\hat{\varepsilon} \neq \hat{\varepsilon}^T$ or $\hat{\mu} \neq \hat{\mu}^T$ [82,86]:

$$\hat{\varepsilon} = \begin{bmatrix} \varepsilon_{xx} & i\alpha & 0 \\ -i\alpha & \varepsilon_{yy} & 0 \\ 0 & 0 & \varepsilon_{zz} \end{bmatrix}, \hat{\mu} = \begin{bmatrix} \mu_{xx} & i\delta & 0 \\ -i\delta & \mu_{yy} & 0 \\ 0 & 0 & \mu_{zz} \end{bmatrix}$$

The quantities $\alpha$ and $\delta$ are gyrotropic parameters. In magnetic materials, they are responsible for the Faraday rotation[87]. In the low-frequency range (radio frequencies (RF), microwaves), there are materials with $\alpha \neq 0$ and $\delta \neq 0$. In the high-frequency range (THz, optics) usually $\hat{\mu} = 1$ and $\delta = 0$. Note that systems with symmetric tensors $\hat{\varepsilon} = \hat{\varepsilon}^T$ or $\hat{\mu} = \hat{\mu}^T$ with nonzero off-diagonal elements can rotate the polarization plane (optical activity) as in the Faraday effect but in a reciprocal way.

For illustration, the S-matrix of ideal nonreciprocal elements, a *gyrator* (nonreciprocal phase accumulation), an *isolator* (unidirectional propagation), and a *three-port circulator* (many-port isolator) are

$$\hat{S}_{gyrator} = \begin{vmatrix} 0 & -1 \\ 1 & 0 \end{vmatrix}, \ \hat{S}_{isolator} = \begin{vmatrix} 0 & 0 \\ 1 & 0 \end{vmatrix}, \ \hat{S}_{3-circulator} = \begin{vmatrix} 0 & 0 & 1 \\ 1 & 0 & 0 \\ 0 & 1 & 0 \end{vmatrix} \tag{5}$$

It worth also mentioning so-called *quasi-circulators* with an S-matrix

$$\hat{S}_{quasi-circulator} = \begin{vmatrix} 0 & 0 & 0 \\ 1 & 0 & 0 \\ 0 & 1 & 0 \end{vmatrix}. \tag{6}$$

Quasi-circulators are bidirectional for one pair of ports [ports 1 and 3], unlike symmetric circulators, where isolation is always unidirectional.

Note that ideal nonreciprocal elements are forbidden in Nature because otherwise, they would violate thermodynamics' second law. Indeed, such nonreciprocal systems would allow heat transfer from a colder object to a hotter one and hence allow the reduction of entropy without thermodynamic work. This situation, called the "thermodynamic paradox," has been resolved by Ishimaru (1962) and Barzilai and Gerosa (1965)[88].

Thus, ideal nonreciprocal elements are forbidden, and *real nonreciprocal systems are always imperfect*. The degree of imperfection is determined by (for any pair of channels i and j) the *return loss*[89]

$$RL = -20\log_{10}(|S_{ii}|), \tag{7}$$

*insertion loss*

$$IL = -20\log_{10}(|S_{ni}|), \tag{8}$$

and *isolation*

$$IX = -20\log_{10}(|S_{ji}|), \tag{9}$$

where port n does not belong to the pair i-j.

So, nonreciprocal systems are always dissipative. *But does the dissipation itself break the time-reversal symmetry?* Indeed, if we consider an electromagnetic field propagation through a macroscopic system (a piece of a lossy dielectric), the wave will get dissipated from the initial energy $I_0$ up to, say, $I_0/2$. However, if we reverse "the electromagnetic part of this picture" in time and launch the wave back, it will dissipate further up to $I_0/4$. Hence, the round trip propagation will not return the system to its initial state ($I_0$) but end it up in another state ($I_0/4$). So, the problem seems to be time-reversal asymmetric and nonreciprocal. However, we know that the transmission coefficients for the left-to-right and right-to-left propagation are identical, and the system is reciprocal. The answer to this paradox is that this is only *an apparent asymmetry* caused by the fact that we did not reverse the "dissipation process" in time. In reality, at the microscale, the dissipation is caused by transforming one sort of particles into another, and if we time-reverse all these transformations involved in the process, the time reverse symmetry will be restored. This exemplifies the point that to reverse a system in time, it is not sufficient to launch a wave back. *One also needs to reverse in time all involved processes*. This is in strong contrast to the case of magnetic bias, where the time reversing of all interactions is not sufficient to reverse the system in time; *one also needs to change the sign of the magnetic field* (to *alter* the system). Similar arguments can be found in Ref.[90].

The time-reversal symmetry is preserved at the quantum single-particle level driven by the Schrödinger equation but has the statistical meaning. Indeed, the change of the time arrow (along with flipping the direction of the magnetic field, if any) allows observing the reverse evolution of the *system's wave function*. However, wave function allows prediction of statistical evaluation of a system (evaluation of an ensemble) but not a single particle. It is easy to come up with a single-particle system that does not obey reciprocity. For example, a photon that hits a dielectric medium from a vacuum (or an electron that hits a space with a certain potential) with dielectric permittivity, say 16 (c-Si in optics), has the probability of passing into the dielectric of 64% and the probability 36% of getting reflected. It is a truly quantum phenomenon as it holds even for *absolutely identical photons*. Thus, a photon that passed through such a dielectric may scatter out from it in the time-reversed scenario making this picture time-reversal asymmetric. The time-reversal symmetry restores only after many experiments.

Thus, reciprocity can be broken in a system with a lack of time-reversal symmetry. The most widespread approach to reciprocity breaking is the biasing of the magnetic field[86] discussed in the next section.

## 2. Nonreciprocity based on magnetic field bias

To begin with, let us consider the nonreciprocity based on magnetic field bias. Magnetic-induced nonreciprocity roots in the Faraday rotation effect[86]. The Faraday effect is the rotation of the plane of linear polarization of light propagating in a medium along a magnetic field. In other words, it is an optical activity induced by a magnetic field. It was discovered by Michael Faraday in 1845. Due to the presence of a magnetic field, the dielectric permittivity depends on the wave polarization, clockwise ($\sigma^+$) or counter-clockwise ($\sigma^-$), and hence the partial waves will propagate with different wavevectors,

$$k_\pm = \frac{\omega}{c}\sqrt{\varepsilon_\pm}, \qquad (10)$$

where $\omega = 2\pi f$, $f$ is the frequency, $c$ is the speed of light, and $\varepsilon_\pm$ is the dielectric permittivity of $\sigma^+$ ($\varepsilon_+$) and $\sigma^-$ ($\varepsilon_-$) polarized light.

If $\text{Re}(k_+) \neq \text{Re}(k_-)$ this is the actual Faraday effect. The wave that undergoing left-to-right propagation acquires the polarization rotation angle $\beta$, Figure 2(a). In the backward propagation (right-to-left), the polarization plane rotates in the same direction giving the total polarization rotation of $2\beta$, Figure 2(a). This is strikingly different from the optical activity effect (chirality) when the round-trip propagation gives the total polarization rotation of 0°. Consider a linearly polarized wave that enters a magnetized medium. Such a wave can be represented as a sum of two circularly polarized waves. As a result, the rotation angle equals $\beta(l) = \frac{k_+ - k_-}{2}l$, where $l$ is the propagation distance. Experimentally, a more convenient way to define $\beta(l)$ consists in the introduction of the *Verdet constant* ($V$) that describes the strength of the Faraday effect for a particular material,

$$\beta(l) = VBl, \qquad (11)$$

where $B$ is the strength of the magnetic field. For the round-trip propagation, the wave acquires the rotation angle of $\beta(2l) = 2\beta(l)$.

However, if $\text{Im}(k_+) \neq \text{Im}(k_-)$, then the waves of different polarization will get different dissipation rate (circular dichroism) and hence the left-to-right and right-to-left waves will differ in amplitude, rather than in phase as in the Faraday effect.

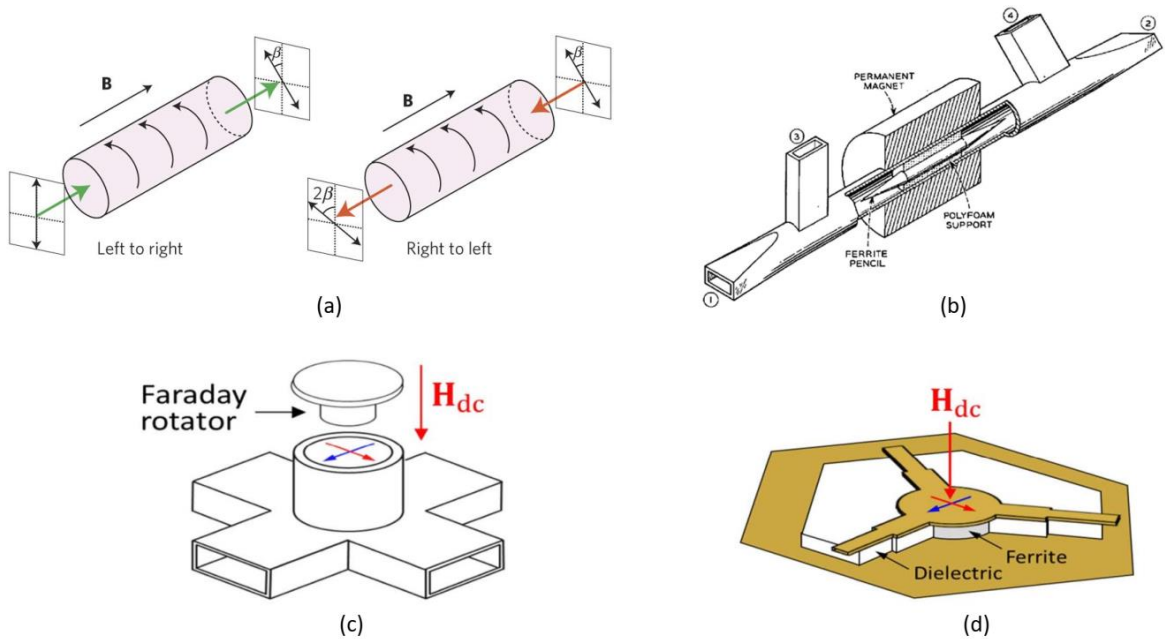

**Figure 2. Nonreciprocity based on magnetic field bias.** (a) Faraday rotation effect. The effect consists in the rotation of a polarization plane caused by the magneto-optical effect, which is the dependence of the wave velocity on the polarization (circular clock- and counterclockwise). The wave that undergoing left-to-right propagation acquires the polarization rotation angle $\beta$. In the backward propagation (right-to-left), the polarization plane rotates in the same direction giving the total polarization rotation of $2\beta$. (b) One of the first Faraday rotation circulators based on a gyrator and tee junctions[91]. (c), (d) Ferrite Y-junction circulators: (c) Waveguide implementation[89], (d) Microstrip implementation[92].

Note that this magnetic-biasing approach works in both the low-frequency: RF and microwaves, and high-frequency: THz[93–95] and optics[96,97]. Employing plasmonic resonance in optics allows boosting the magneto-optical effects[98–100].

In most materials, a high Verdet constant is measured near the frequencies of optical transitions in atoms where strong absorption occurs. The typical values of the Verdet constant [ rad/(T×m) ] are: $V = 60-475$ for $Tb_3Ga_5O_{12}$ (TGG) at 333–750 THz[101,102], $V = 384-5760$ for Bi-doped YIG at 385–553 THz[103], $V = 60-475$ for Rb vapor at 385 THz[104], and at the order of few thousands in YbBi:YIG at 170–300 THz[105]. For Verdet constants of other magneto-active materials, we refer to Ref.[106].

An optical isolator can be realized when a Faraday medium is positioned between two polarizers set at π/4, with an induced rotation of π/4. This arrangement provides high transmission in one direction and isolation in the other[96,104].

Figure 2(b) demonstrates one of the first Faraday rotation circulators proposed by Fox et al. in 1955 [91]. The operation principle is based on combining a gyrator with power dividers or couplers. For example, if the wave enters port#1, it does not couple to the port#3 as they are orthogonal; then, the gyrator rotates polarization so that the wave can now effectively couple to port#4, leaving port#2 uncoupled as well. However, because of the Faraday rotation effect, the wave that enters port#4 does not go to the port#1, but effectively couples to the port#3. One can continue this consideration to check that this system is fully nonreciprocal.

Figures 2(c) and (d) demonstrate the configurations of ferrite Y-junction circulators that are popular today. These devices contain the ferrite magnetic disk and differ only in the way of realization: (a) waveguide[89] and (b) microstrip[92] geometry. In these devices, without external magnetic bias, the electrons' magnetic dipoles inside the ferrite disk are randomly oriented, and hence the counterrotating modes are degenerate. In this unbiased regime, the junction operates as a regular divider. When the external magnetic bias is applied, the spinning electrons' dipole moments align in the same direction. Because of Eq.(10), the mode propagating in the direction of the magnetic precession will exhibit a different propagation velocity than the one propagating in the opposite direction[107].

## 3. Advanced nonreciprocal materials: Weyl Semimetals, Metamaterials, magnetic 2D materials, topological insulators

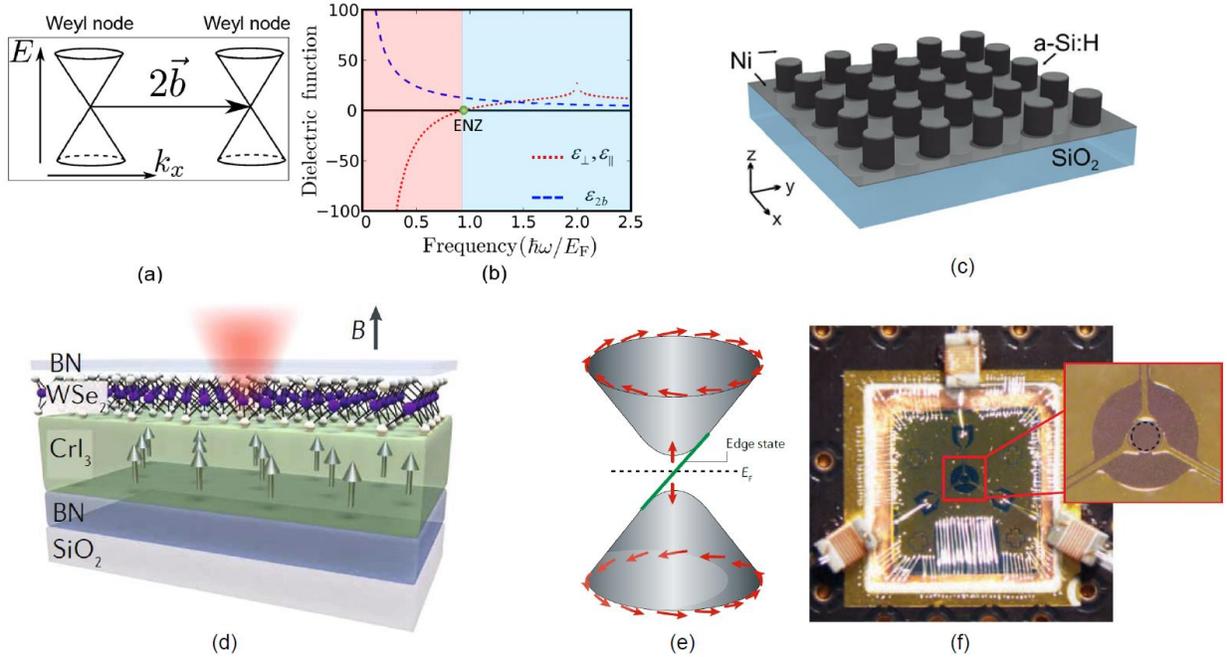

**Figure 3. Novel nonreciprocal materials.** (a,b) Weyl semimetals. (a) Weyl nodes in Weyl semimetals. (b) Dispersion of the dielectric functions of an isotropic Weyl semimetal. ENZ stands for the epsilon-near-zero regime. The red area corresponds to the metallic phase, blue – dielectric. (c) Schematic illustration of the magnetophotonic metasurface composed of Si nanodisks supporting magnetic Mie-type resonances. Disks are covered by a 5-nm-thick Ni film[108]. (d) magnetic 2D materials. Magnetic proximity coupling between monolayer $WSe_2$ and 2D $CrI_3$ induces valley splitting in the $WSe_2$ layer[109]. (e) The gapped Dirac-like dispersion of the surface state in a magnetic topological insulator[110]. (f) Photograph of quantum Hall circulator device showing the three coplanar transmission lines connected to copper wire wound chip inductors for impedance matching. Inset: Photo of the circulator showing a 330-μm diameter 2DEG disc with a 20-μm gap to the metal defining the three signal ports[111].

In order to enhance the intrinsic magneto-optical effects of natural materials, it has been suggested to utilize the ability of resonant plasmonic and dielectric nanostructures to localize light and enhance light-matter interaction[112–114]. This is the essence of so-called Nanoscale magnetophotonics[114]. For example, a combination of noble and magnetic materials in nanoparticles[115], embedding gold nanoparticles in magnetic dielectric[116,117], and ferromagnetic nanodisks[118,119] have been demonstrated. The enhancement of magneto-optical effects can be reached via the excitation of surface plasmon-polaritons[120,121] and Mie resonances[122,123]. Tailoring electromagnetic systems with unusual scattering effects like bound states in the continuum (BICs) with unboundedly large Q-factor[124–127] is another route to efficient nonreciprocal structures and devices[128,129].

Further investigation in the area of nonreciprocal devices based on magnetic field bias is mainly focused on new materials such as *Weyl Semimetals*[130–133] (Figure 3(a,b)), *metamaterials and metasurfaces*[108,134] (Figure 3(c)), *magnetic 2D materials and heterostructures*[135] (Figure 3(d)) and *topological insulators*[136,137] (Figure 3(e,f)). In this section, we briefly overview these important research areas with a focus on their nonreciprocal response.

A recent revolution in material science and condensed matter physics has started with discovering *natural topological insulators* with the nontrivial topology of their electronic band structure[138–142]. A unique ingredient for turning such materials into topological insulators is the presence of the spin-orbit coupling. When significant spin-orbit interaction is present, a bandgap opens at the bands' overlap region, and the materials turn into topological insulators. As a result, topological insulators possess a bandgap in the band structure like an ordinary insulator and gapless surface states sustained by an odd number of 2D Dirac fermions with helical spin texture protected by the topology [138]. These topologically protected surface states exhibit the remarkable property of spin-locked propagation of currents with forbidden back-scatterings, which is of great interest for electronics, optoelectronics, spintronics, and quantum computation. Analogous concepts have been recently realized in photonics, giving rise to topological photonic insulators [9,143].

Further works have demonstrated that the edge-bulk correspondence can also be implemented when the bulk is gapless in so-called bulk Dirac Semimetals (DSs). DSs have been predicted and demonstrated as "3D graphene" states of matter. Although 3D Dirac states in DSs are not topologically protected as 2D Dirac states on topological insulators' surface, they have crystalline symmetry protection against gap formation similar to ordinary graphene [144].

Topologically nontrivial and robust band massless *3D Dirac states require time-reversal or inversion symmetry breaking* (or both) to lift the degeneracy of 3D Dirac band-touching points. These states have been recently discovered in topological *Weyl semimetals* (WS) – a new topologically nontrivial phase of matter – where the lack of symmetry mentioned above leads to massless bulk fermions and topologically protected Fermi arc surface states [132]. The Weyl semimetals' (like $WTe_2$, $TaIrTe_4$, $NbP$, $Co_2$-based Heusler compounds) band structure embraces an even number of nondegenerate band-touching points arising at the Fermi level (*Weyl nodes*) and split in the momentum space [Figure 3(a)], which gives rise to their topological stability. The Weyl points appear in pairs, exhibit a linear dispersion around them, and can have positive or negative helicity charges. The existence of such defined helicity charge unites the Weyl points with helicity-degenerate valleys in 2D Transition Metal Dichalcogenides (TMDC), which recently emerged as a versatile platform for valleytronics [145]. The superior properties of Weyl semimetals in microwave and THz frequency ranges, including relatively low loss caused by a large momentum scattering time ~1 ps at room temperature[146], epsilon-near-zero and hyperbolic regime, make them very promising for electronic and optoelectronic applications.

Weyl semimetals' optical properties have been studied theoretically in a number of papers, e.g., Refs. [133,147]. In these works, it has been demonstrated that to account for WS topological properties in the optical response, one may use the standard form of Maxwell equations $\mathbf{D} = \hat{\varepsilon}_{WS}\mathbf{E}$ taking the WS (relative) permittivity tensor (and unity magnetic permeability) in the form [133,147]

$$\hat{\varepsilon}_{WS} = \begin{pmatrix} \varepsilon_\perp & i\varepsilon_{2b} & 0 \\ -i\varepsilon_{2b} & \varepsilon_\perp & 0 \\ 0 & 0 & \varepsilon_\parallel \end{pmatrix}, \quad \varepsilon_{2b} = \frac{e^2}{\pi\hbar\omega} 2|\mathbf{b}| \quad (12)$$

where $\varepsilon_\perp$, $\varepsilon_\parallel$ are diagonal elements, and $\varepsilon_{2b}$ is the nondiagonal component responsible for the strength of the media's magneto-optical activity, here $e$ is the charge of electrons, $\omega$ is the angular frequency. *This component is caused by the Weyl nodes splitting in the momentum space by the vector* 2**b**. The longitudinal dielectric tensor component has the Drude form $\varepsilon_\parallel = \varepsilon_h - \omega_p^2 / (\omega^2 - q_z^2 v_F^2)$, where $\varepsilon_h$ is the background dielectric constant, $\omega_p$ is the plasma frequency, $v_F$ is the Fermi velocity, and $q_z$ is the longitudinal component of the wavevector[148].

For the frequencies far away from the plasma resonance ($\omega \gg \omega_p$), the typical dispersion of the isotropic WSs permittivity (assuming $\varepsilon_\perp = \varepsilon_\parallel$) is presented in Figure 3(b). The permittivity has a negative real part at low frequencies $\hbar\omega < E_F$, an epsilon-near-zero (ENZ) region ($\hbar\omega \approx E_F$), and a high-index dielectric regime ($\hbar\omega > E_F$). Here, $E_F$ stands for the Fermi energy. Hence, WSs provide a versatile platform for realizing a plethora of promising optical effects, with the unique advantage of strong interactions with its electronic properties and an ultrathin, planarized platform.

In the isotropic case ($\varepsilon_\perp = \varepsilon_\parallel$), Weyl semimetals support traditional surface plasmon polariton (SPP) waves at $\hbar\omega < E_F$ on the interface with a dielectric, i.e., when the diagonal permittivity is negative. These waves have been lately discussed theoretically, and their ability to carry energy in one single direction (nonreciprocity) has been predicted [133]. A similar approach has recently been suggested for topologically protected one-way waveguiding in the 3D scenario in microwave and THz frequency ranges[149]. In general, the Weyl semimetal-based structures' topological robustness enables reflectionless routing of electric currents and electromagnetic waves along arbitrary pathways, making them promising for a novel generation of integrated optoelectronic and photonic devices which also can be made robust against disorder.

In Ref.[130] the Faraday isolator based on the Weyl semimetal slab has been suggested. In the mid-IR frequency range (λ = 3.5 μm), the isolation of 41.3 dB supported by a shallow insertion loss of 0.33 dB has been reported. Notably, such great isolation property occurs for the Weyl slab of the exceptionally small thickness of λ/4. It is shown that the isolation characteristics can be further improved in the geometry of a photonic crystal.

In the realm of light absorption and (thermal) emission, the reciprocity principle is expressed by Kirchhoff's law of thermal radiation that dictates that for reciprocal systems, the emissivity and absorptivity are restricted to be equal[83,150]. In recent work, however, it has been reported that *nonreciprocal thermal emitters* can be realized using topological magnetic Weyl semimetals[131] that don't require strong magnetic fields (~0.3 - 3T) as in other existing approaches based on conventional magneto-optical effects[151,152]. The authors predict Kirchhoff's law's violation in a broad angular and frequency range and its high temperature-sensitivity.

Optically dense, intelligently designed structures composed of electromagnetically small nanoelements periodically or quasi-periodically arranged on a metal or dielectric substrate with subwavelength relative distances have been conventionally called *metasurfaces*[8,153–159]. Investigation of these flat materials takes its origin in the area of metamaterials, which is a more general concept of artificial bulk media, allowing tailoring optical, acoustical, or mechanical properties on demand[160]. Metamaterials' appearance has also allowed revisiting the fundamental concepts of wave physics, which resulted in an appearance of astounding novel effects, for example, negative refraction, backward waves, surpassing the diffraction limit for subwavelength imaging[161–163]. Recently, it has been understood that many functionalities offered by metamaterials can be realized with thin layers of bulk metamaterials or even with their 2D layouts – metasurfaces. The ultimate 2D case caused a particular interest due to the fabrication simplicity and optical fineness, which relaxes many undesirable limitations such as bulk absorption and phase-matching in nonlinear photonics. Although such flat electromagnetic structures with subwavelength granularity were known in micro- and radio-waves since the $1950^{th}$ as transmission array antennas (e.g.[164]) and frequency-selective surfaces[154,155], they experienced an explosive rebirth owing to developments in nanofabrication techniques that have allowed the creation of such structures in the optical range.

Initially, in the microwave and optical frequency ranges, metamaterials and metasurfaces have been designed using subwavelength metallic elements [153,165–167]. The recent interest is concerned with the design of dielectric metamaterials and metasurfaces and the investigation of their properties in the optical frequency range[168–172]. Using dielectric materials such as silicon reduces ohmic losses compared to classical plasmonic structures as well as tailoring both the electrical and magnetic components of light, thereby increasing the overall efficiency in subwavelength structures. For example, current trends favor dielectric metalenses instead of initially proposed metals in photonics[173,174].

The locally engineered properties of optical surface conductivity allow us to realize conventional and exotic functionality, including anomalous reflection and refraction, polarization conversion, enhanced light emission, sensing, imaging, and communications, with optically thin surfaces, making metasurfaces a new paradigm in optics. Their subsequent integration with light-emitting and harvesting materials and coming up with novel functionalities like topology, PT-symmetry, embedded eigenstates, etc. paved the way for the metasurfaces application well beyond wave- and classical optics. It has been recently demonstrated that their compact dimensions along with the unique ability to tailoring near-field distribution and local density of states, the concept of geometric phase[175–177], enhanced nonlinearity[178] and light-matter interactions, unusual scattering effects[179], make metasurfaces an emerging platform for quantum optics applications.

Metasurfaces offer promising opportunities for magneto-optical and nonreciprocal effects. For example, in Ref. [108] it has been shown that the multifold enhancement of the magneto-optical response both in Faraday and Voigt configurations in the spectral vicinity of the magnetic dipole

Mie resonances of Si nanoparticles covered with a thin magnetic film (Ni) can be achieved, Figure 3(c). The Faraday configuration consists of a transparent slab magnetized along the wave propagation encompassed by two linear polarizers twisted at 45°. In the Voigt configuration, the external magnetic field or magnetic polarisation is applied along the slab perpendicular to the light wave vector. The proposed Si metasurface can be placed atop of a magnetically-biased material like in Ref.[180]. In another work[181], the ability of gold (Au) metasurface introduced into a thin magnetic garnet film to the accurate tuning of the transverse magneto-optical Kerr effect was experimentally demonstrated. Another perspective approach is to utilize bimetallic metaatoms[182]. The other types of nonreciprocal but nongyrotropic non-magnetic metasurfaces are discussed in what follows[32,80,183–187].

Provoked by the recent discovery of magnetic 2D materials, the discussed magneto-optical effects, the nonreciprocal response, and the devices exploiting them go to nanoscale and even to the atomic scale[109,135,188]. The examples of magnetic materials that do exist in the 2D phase and attract a lot of interest are $CrI_3$[189], $Fe_3GeTe_2$[190], $CrGeTe_3$, $Cr_2Ge_2Te_6$[191], and $FePS_3$[135]. An extensive review of magneto-optic effects in 2D materials and related applications can be found in Ref.[192]. Note that magnetic 2D materials are used either by themselves for a magneto-optical response or in pair with other 2D or quasi 2D materials. For example, Figure 3(d) demonstrates a heterostructure formed by an ultrathin ferromagnetic semiconductor $CrI_3$ and a monolayer of $WSe_2$. Unprecedented control of the spin and valley pseudospin in $WSe_2$ in this work has been observed. The authors have demonstrated a magnetic exchange field of nearly 13T and rapid switching of the $WSe_2$ valley splitting and polarization via flipping of the $CrI_3$ magnetization[109]. Since valley polarization in transition metal dichalcogenides attracts more and more attention for nonreciprocity [193], their combination with magnetic 2D materials will be very beneficial for nonreciprocal devices. It worth noting that the Faraday effect in graphene has also been intensively investigated[194,195]. It has been demonstrated that under a 7T magnetic field, single-layer graphene can induce a large Faraday rotation with a rotated angle of 0.1 rad. The Faraday effect in magnetic 2D materials can be further enhanced by assembling heterostructures[196], periodicity[197], and introducing plasmon coupling[98,198].

Finally, as stressed above, topological insulators are of great interest and importance for nonreciprocity and time-reversal symmetry breaking[110]. Usually, topological insulators have Dirac-like dispersion caused by the honey-comb arrangement of atoms. If the material itself is 2D, as in graphene, then the *quantum Hall effect* can arise[199] when a magnetic field is applied orthogonally to the surface of graphene. This is accompanied by a splitting of the Dirac cone with the formation of the gap and surface states that connect the valence and conductive bands of graphene. Now the current is enforced to propagate only along the edges. Remarkably, these edge states are unidirectional, scatter-less and robust to the presence of defects on edge. This effect is called the quantum Hall effect. It was the first topological effect to be discovered in the 1980s. Another topological phase of matter, the *quantum spin Hall effect*, does not require magnetic field bias, was theoretically predicted in 2006 and experimentally observed in 2007 in HgTe quantum wells. This effect relies on the strong spin-orbit coupling that leads to spin-momentum locking[200].

Much more recently, the 3D topological phases have also been observed, for example, in $Bi_{1-x}Sb_x$, $Bi_2Se_3$, $Bi_2Te_3$, and $Sb_2Te_3$, and now are of enormous interest for electronics and spintronics[138]. Similar to 2D systems, 2D topological insulators are insulators in their volume but can conduct scatter-less and disorder-tolerant currents on their surface. They also do not require the presence of the magnetic field because the gapped Dirac-like dispersion of the surface state in a magnetic topological insulator comes from the strong spin-orbit coupling[110], Figure 3(e).

Figure 3(f) demonstrates the realized on-chip microwave circulator based on the nonreciprocal response caused by the quantum Hall effect. The inset shows the photo of the circulator with a 330-μm diameter disc of two-dimensional electron gas (interface of the semiconductors GaAs and AlGaAs) with a 20-μm gap to the metal defining the three signal ports[111]. The work reports a *dynamically tuned nonreciprocity* of 25 dB over a 50-MHz bandwidth. The theoretical analysis of this system can be found in Ref.[201]. Note that such a device can also be realized without magnetic field bias using topological insulator devices that exhibit the quantum anomalous Hall effect[110].

## 4. Active nonreciprocity

While non-Hermitian gainy systems by themselves cannot break reciprocity, they can support a *unidirectional* operation regime giving rise to the active nonreciprocity mechanisms. This type of reciprocity braking relies on the fact that DC biased transistors and transistor amplifiers exhibit unidirectional gain and, as such, are two-port nonreciprocal devices[202]. The first of such devices were proposed in the early 1960s and were gyrators based on *negative-impedance converters* (NICs) [203] or *operational amplifiers*[204,205], and have been actively using in the full-duplex systems for the simultaneous transmission and reception of signals by the same antenna at the same frequency.

The first active circulator was proposed by Tanaka et al. [206] in 1965. The device consisted of three transistors, sequentially connected in a loop, as shown in Figure 4(a). Tanaka's circulator does not involve reactive components (active components used for matching), and as such, it is *very wideband*[206]. Of course, in practice, the bandwidth is limited by the parasitic capacitances of the transistors.

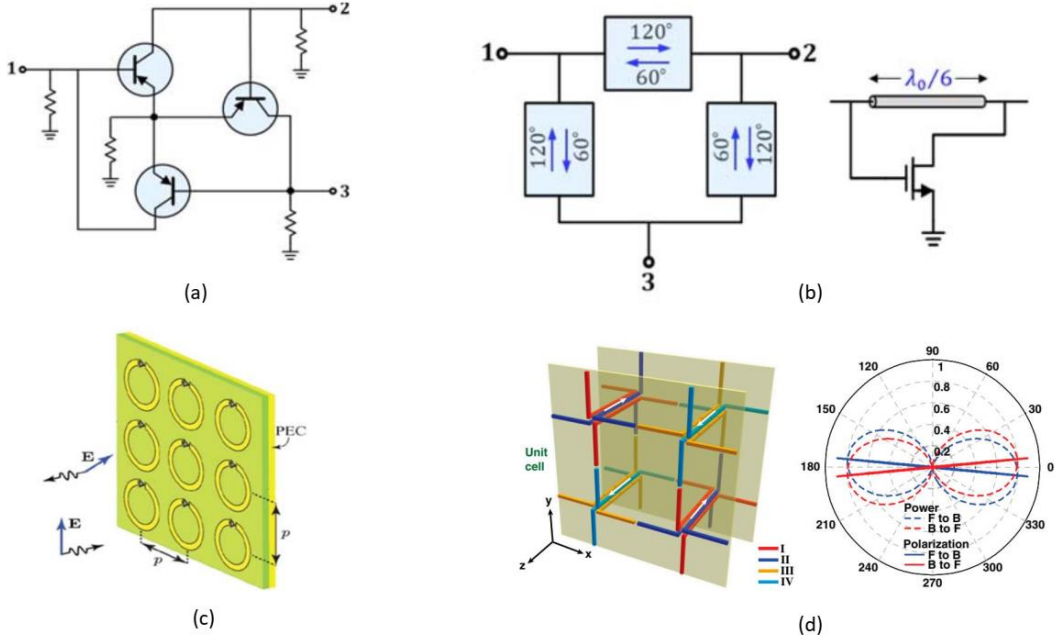

**Figure 4**. **Active nonreciprocity in RF and microwaves.** Active nonreciprocity in radiofrequency and microwaves. (a) Tanaka's architecture[206]. (b) Ayasli's architecture[207]. (c,d) Active nonreciprocal metamaterials: (c) metasurface based on an array of transistor-loaded microstrip ring resonators proposed by Kodera et al.[208]. (d) Left: The architecture proposed by Wang et al.[209] using two antenna arrays with orthogonal polarizations and unidirectional coupling through amplifiers. Right: nonreciprocal rotation of the polarization state for front-to-back and back-to-front propagation.

Ayasli implemented nonreciprocal phase shifters as a parallel combination of transmission lines and transistors, Figure 4(b), but a lumped element implementation is also possible[210]. The transmission lines are designed to exhibit a bidirectional phase delay of 60°, while transistors transmit signals unidirectionally with a phase delay of 180°. As in magnetic circulators, isolation is achieved through destructive interference of signals reaching a certain port from opposite directions. Due to the phase shifters' dispersion, this condition is perfectly satisfied only at a single frequency, making this circulator relatively narrowband.

Among these realizations, Tanaka's approach provides the best performance in terms of power handling and noise. However, as Carchon and Nanwelaers[211] have shown, this approach *still lags far behind ferrite and parametric circulators*.

Beyond circulators, transistors' unidirectional properties have been explored in the design of nonreciprocal metamaterials and metasurfaces. Kodera et al. proposed one such metamaterial[212] based on an array of transistor-loaded split-ring resonators, as shown in Figure 4(c). A split-ring resonator supports two counterpropagating modes with the same resonance frequency, which are degenerated due to the reciprocity. The addition of the transistor breaks one of these modes and makes the ring resonant for only one polarization, similar to magnetic-biased ferrites[202]. In subsequent articles, Kodera et al. have also demonstrated that this metamaterial-based approach can be used for integrated nonreciprocal devices, such as circulators, isolators, and nonreciprocal antennas[208,213].

In Ref. [209], Wang et al. have suggested a similar concept using two antenna arrays with orthogonal polarizations and unidirectional coupling through amplifiers, as shown in Figure 4(d). A similar structure was proposed by Popa and Cummer [214] based on two arrays of electric and magnetic dipoles connected through an array of unidirectional amplifiers. As a result, this metamaterial provides nonreciprocal rotation of the polarization state for front-to-back and back-to-front propagation, Figure 4(d, left).

In the subsequent works, this idea of actively loaded metasurfaces has been actively exploited for designing new nonreciprocal devices, including magnet-free Tellegen and moving metasurfaces[185], surface-circuit-surface unidirectional metasurfaces[183], and space-energy digital-coding metasurface[215].

Another interesting regime attracting attention for nonreciprocal optical systems enabled in non-Hernissian systems with gain is the so-called parity-time-symmetry (PT-symmetry) phase [43,216,217]. This regime can be illustrated in a system of two coupled resonators with resonant frequencies $\omega_1$ and $\omega_2$, and decay rates $+\gamma_1$ (loss) and $-\gamma_2 = g$ (gain). The modes are coupled with the coupling strength $\kappa$. The analysis of this system shows that, if the coupling is weaker than a certain critical value ($\kappa < \kappa_{PT}$), the system possesses two modes, one lossy and one amplifying. When excited, the lossy mode decays exponentially in time, whereas the other exhibits exponential growth. However, if the coupling strength is large enough ($\kappa > \kappa_{PT}$), the system resides in the strong coupling regime whereby coherent energy exchange between resonators compensates for the loss decay, stabilizing the system in the PT-symmetric phase. The critical point $\kappa = \kappa_{PT}$ is called an *exceptional point* (EP) because it corresponds to the coalescence of two poles at the real frequency axis[85,218–223]. In the PT-symmetry broken phase, the system can exhibit unidirectional gain amplifying waves that enter the system from a certain port and attenuate the waves in other ports.

The idea of using systems with PT-symmetry for breaking reciprocity has been suggested in Refs. [43,216]. In these works, the PT-symmetrical phase for optical isolation has been experimentally realized on a chip at the wavelength of ~1,550 nm using two directly coupled high-Q silica microdisk resonators. Good isolating properties have been measured in both systems, stemming from the strong asymmetry along with highly localized fields and hence strong nonlinearity in the PT-symmetry broken phase. As a result, the wave that comes from one port gets dissipated, but the one that comes from the opposite port gets amplified and nonreciprocally transmitted out. These ideas have recently been further developed in Refs.[224–228].

In conclusion, it worth noting that this approach is not free of drawbacks. Although active nonreciprocal components are attractive for their compatibility with on-chip implementation and relatively broad bandwidth, they suffer from a low power handling and high noise. The realized circulators based on active elements do not outperform the magnetic-based nonreciprocal circulators. The recent manifestation of the so-called virtual gain effect[229,230] holds a promise of passive nonlinear nonreciprocal systems free of the mentioned drawbacks.

## 5. Nonreciprocity based on time-modulation

In 2013 D. Sounas et al. have proposed a new approach to achieve magnetic-free nonreciprocity via angular momentum biasing of resonant rings with specifically tailored spatiotemporal azimuthal modulation[80], Figure 5(a). It has been demonstrated that the proposed form of modulation removes the degeneracy between opposite resonant states of a resonator that can then be used as a meta-atom in metamaterials and metasurfaces. In combination with a suitably induced high-Q response, the approach provides a giant nonreciprocity on a subwavelength footprint. A few applications based on this approach have been subsequently proposed, including an ultrathin RF isolator, Faraday rotation, and an optical isolator, all realized without requiring bulky magnetic biasing elements[19]. This approach is currently known as *spatiotemporal modulation biasing*.

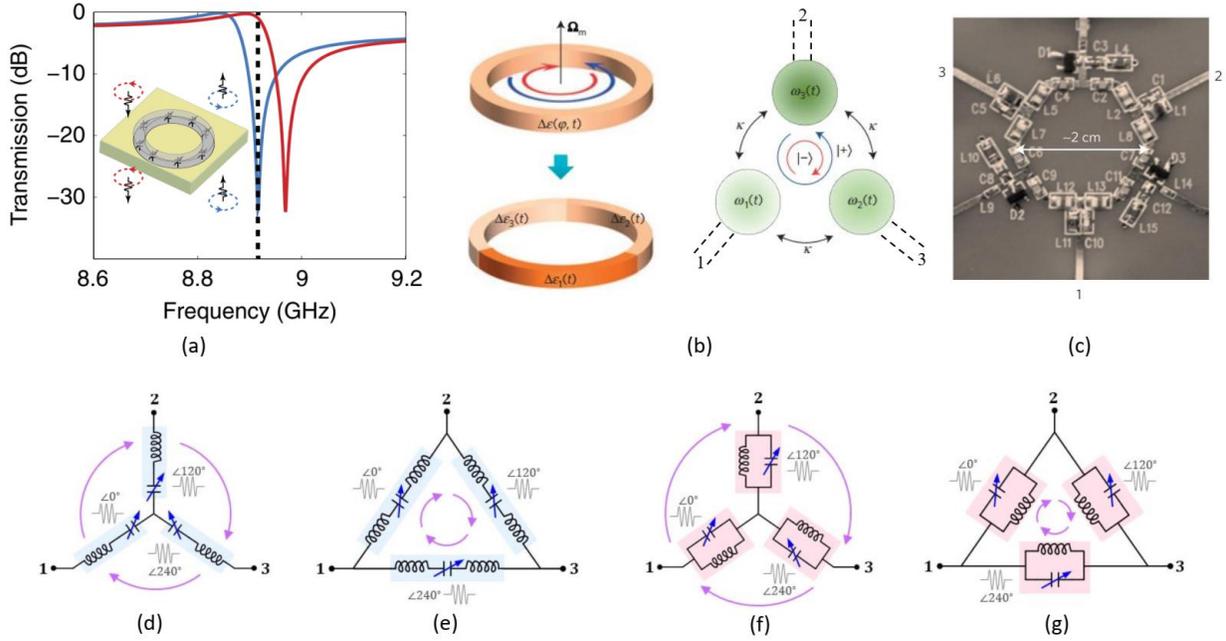

**Figure 5. Nonreciprocity based on time-modulation.** Magnetic-free nonreciprocity with synthetic angular-momentum biasing. (a) Transmission across the metasurface with the time-modulated (via an array of varactor diodes) unit cell[80]. (b) Left: An azimuthally symmetric ring with spatiotemporal permittivity modulation. The modulation imparts an effective electronic spin to the ring, with a certain angular velocity, which lifts the degeneracy of counterpropagating resonant states and induces nonreciprocity. Its practical realization typically involves discretizing into N different regions. Right: The topology of a loop consisting of three strongly coupled identical resonators, modulated with a 120º phase-difference. (c) Fabricated prototype. The maximum dimension of the structure is 2 cm, corresponding to an electrical size of λ/75 at 200

MHz[21]. (d - g) Single-ended STM-AM circulator topologies: (d) Bandpass/wye [213]. (e) Bandstop/delta [231]. (f) Bandstop/wye [231]. (g) Bandpass/delta [84,231].

Inspired by this approach of explicit time modulation, Estep et al.[21] realized a magnetless circulator based on three coupled resonators with a variable resonant frequency. Figure 5(b) illustrates the conceptual diagram of how the concept works. Here three resonators are mutually coupled to form a resonant loop. The natural oscillation frequencies ($\omega_0$) of the resonators are modulated with signals having 0°, 120°, and 240° phases

$$\omega_n = \omega_0 + \Delta\omega \cos\left[\omega_m t + (n-1)\frac{2\pi}{3}\right], \tag{13}$$

where n = 1, 2, 3 is the resonators' number, $\omega$ and $\omega_m$ are the modulation depth and rate, respectively. As a result, the clockwise and counterclockwise modes acquire different phases, which can then be used for isolation and circulation. Figure 5(c) demonstrates the fabricated prototype. The maximum dimension of the structure was 2 cm, corresponding to an electrical size of λ/75 at 200 MHz. Figure 5(d - g) illustrates the single-ended STM-AM circulator topologies: (d) Bandpass/wye [213]. (e) Bandstop/delta [231]. (f) Bandstop/wye [231]. (g) Bandpass/delta [84,231].

It has been demonstrated that the single spatiotemporal modulation circulator outperforms the previous works in terms of power handling and noise figure by several orders of magnitude. However, this kind of circulator *generally suffers from high intermodulation products* (IMPs) resulting from mixing between the input and modulation signals. The intermodulation is caused either by the time-varying nature of the circuits or the inevitable nonlinearity. These IMPs convert power from the fundamental harmonic to other frequencies, limiting the lowest possible IL to ~3 dB.

In order to get around this drawback, a more effective solution based on combining two single-ended circuits in a differential voltage- or current-mode architecture and maintaining a constant 180° phase difference between their modulation signals has been developed[232]. Also, N-way spatiotemporal modulation circulators have been suggested in Refs.[231,233]. This is done by dividing the input power into equal portions, each transmitted through a single circulator, and then summing up them again at the output port[81]. With the growth of N, the N-way spatiotemporal modulation circulators become more immune to any practical nonidealities. However, this comes with the price of complexity and power consumption.

Another advantage of this spatiotemporal modulation approach is its *relatively wide bandwidth*. For example, a broadband spatiotemporal modulation circulator based on a current-mode bandpass/wye junction and the second-order matching networks was experimentally validated in [234]. Such a circuit's bandwidth was reported to be 14% (140 MHz), which is more than sufficient for all potential applications.

We note that for the sake of compactness, in RF and microwaves, it is more convenient to use switched capacitors rather than varactors to achieve the modulation in any of the previous spatiotemporal modulation circuits. This is mainly because varactors typically require complex networks with multiple large inductors to inject the modulation signals into the junction and prohibit their leakage to the circulator's RF ports. The first integrated-circuit implementation of spatiotemporal modulation circulators, based on a modified version of the current-mode bandpass/wye architecture, was presented in Ref.[235] at 910 MHz using a 180-nm CMOS process. Another example based on the single-ended bandstop/delta topology was also presented at ~1 GHz in [236].

The main drawback of the time-modulation approach illustrated in Figure 5 is its bad scalability to high frequencies. Recently, however, several systems based on the time modulation of conductivity of graphene have been proposed at THz and even IR threesome ranges[237,238]. Although this approach is not realized yet, time modulation of graphene appears as a compelling platform for nonreciprocal devices on a chip due to its remarkable optoelectronic properties and surface plasmon modes.

Other promising platforms for nonreciprocal devices that can overcome those based on gyrotropic materials at terahertz, IR, and telecom wavelengths are *p-i-n* junctions[22] and narrow-band optomechanical effects[26,39,40,239–241].

## 6. Nonreciprocity based on nonlinearity

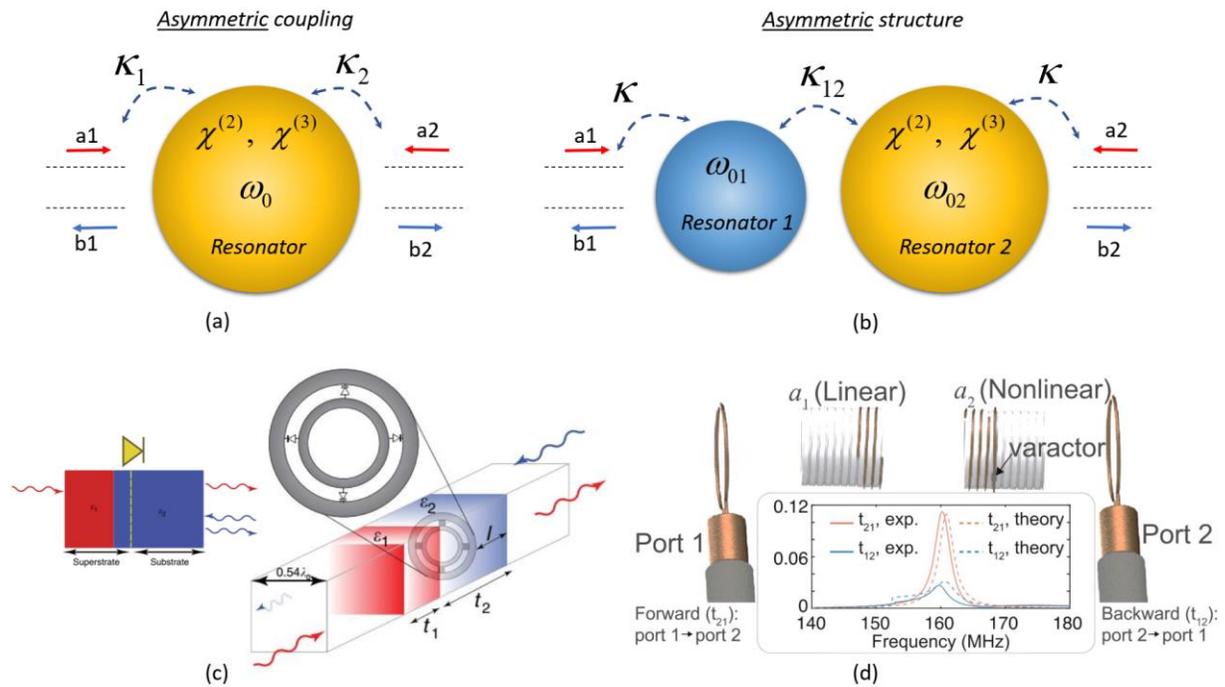

**Figure 6**. **Nonreciprocity based on nonlinearity**. Two kinds of nonlinearity-asymmetry-based structures: (a) a nonlinear resonator (cavity) with an asymmetric coupling to the channels and (b) essentially asymmetric nonlinear structure of two or more resonators. (c) Example illustrating the approach (a). Two dielectrics of different dielectric permittivity provide an asymmetry in the electric field distribution under left- and rightward excitation. The inset shows the varactor-loaded thin resonator (two concentric rings) placed at some distance from one of the slab ends. The bilayered slab and the ring resonator are placed within a rectangular waveguide[186]. (d) Work that illustrates the approach (b). The system consists of two coupled resonant coils; one of those is nonlinear due to a varactor[242].

The next approach to nonreciprocal behavior we discuss is based on nonlinearity in spatially asymmetric structures. Figures 6(a) and (b) illustrate two possible scenarios. The first one [Figure 6(a)] relies on a nonlinear (that is, the resonant frequency, $\omega_0$, depends on the energy in the cavity) cavity asymmetrically coupled to the excitation channels ($\kappa_1 \neq \kappa_2$). In symmetric coupling $\kappa_1 = \kappa_2$, the system is obviously reciprocal, no matter how strong the nonlinearity is. The second scenario [Figure 6(b)] is formed by at least two coupled cavities, where at least one is nonlinear.

The work [186] illustrates the approach of Figure 6(a). Here, two dielectrics of different dielectric permittivity provide an asymmetry in the electric field distribution under left- and rightward excitation, Figure 6(c). The inset in Figure 6(c) shows the varactor-loaded thin resonator (two concentric rings) placed at some distance from one of the slab ends. The bilayered slab and the ring resonator can be placed within a rectangular waveguide giving rise to a nonreciprocal 1D system. This structure can also be used as a unit cell for a nonreciprocal metasurface or metamaterial[186].

This approach can be realized in the optical realm utilizing different kinds of nonlinearity, including Kerr-nonlinearity, liquid crystals, phase-change materials, electron-hole plasma generation, etc. Among these, the one based on phase-change materials attracts increasing interest today due to strong nonlinear response and relatively fast response[243,244]. In Ref. [245], an optical diode based on asymmetric optical absorption in a vanadium dioxide ($VO_2$) film embedded in a simple thin-film assembly has been reported. The large change in complex refractive index across the insulator-to-metal transition in $VO_2$ makes it possible to design ultrathin limiting optical diodes. The reported device features a nonreciprocal high-power-state transmission of ~0.22 for the forward incidence and ~0.04 for backward incidence. This simple device is broadband, ten times thinner than the operating wavelength, and can be useful for the protection of laser sources from parasitic back-scattering.

This nonlinearity can be realized in the microwave range with a varactor diode coupled to a resonator, for example, a split-ring resonator. The varactor diode is a nonlinear element actively used in photonics and metamaterials[246,247]. For example, in Ref.[246] dielectric discs supporting Mie resonances coupled to varactor diodes have been investigated in order to achieve *topologically protected edge states* at frequencies around 1.5 GHz.

In the quantum regime on a few-photon level, the geometry illustrated in Figure 6(a) with the coupling rates' asymmetry has been suggested and realized in Ref.[64]. In this work, a cavity is filled with two-level atoms supporting quantum nonlinearity (population saturation of the excited state). The atoms have provided the nonlinearity, whereas the difference in the cavity's walls' transmission coefficients yelled the required asymmetry.

The recent work [242] illustrates the second approach [see Figure 6(b)] in the microwave frequency range. The system is based on two coupled linear-nonlinear resonators, Figure 5(d). The theoretical analysis of this general coupled system gives isolation [$20\log(|t_{21}|/|t_{12}|)$] of ~10 dB. The experimentally realized system consists of two coupled resonant coils supporting magnetic resonance. One of the resonant coils is nonlinear due to the varactor diode. The experimental realization demonstrates that the nonreciprocity depends on the input power and that the maximum forward-backward isolation contrast yields ~20 dB for the input power of 10 dBm.

It worth noting that the structure of asymmetric nonlinear coupled resonators [Figure 6(b)] is more beneficial in terms of design and non-reciprocal performance because it allows tailoring the structural "asymmetry" by the mutual coupling ($\kappa_{12}$) between the cavities.

Importantly, it can be rigorously shown that there is a *fundamental trade-off* between the maximum forward transmission ($T_{fw}$) of maximal transmission and the range of powers over which nonreciprocal transmission can occur[248–250]. Mathematically it can be expressed in the following way

$$T_{fw} \leq \frac{4 \cdot \eta}{(\eta+1)^2}. \tag{14}$$

where $\eta$ is the asymmetry factor defined as $\eta = |\mathbf{E}_2|^2/|\mathbf{E}_1|^2 \geq 1$, and $\mathbf{E}_2$ ($\mathbf{E}_1$) is the field exiting the structure from the first (second) direction. In symmetric structures, $\eta = 1$. The only way to increase transmission consists in the reduction of $\eta$, but it comes with the price of reduction of the nonreciprocity effect[248]. This trade-off works for a very general category of nonlinear isolators. Namely, any structure consisting of one nonlinear and N-1 linear resonators obey this trade-off regardless of the type of the resonance (e.g., Lorentzian or Fano).

However, it was demonstrated that this trade-off could be overcome by utilizing more nonlinear resonators. For instance, in Ref.[250] it was shown that a combination of one Fano and one Lorentzian nonlinear resonator, and a suitable delay line between them, can provide unitary transmission ($T_{fw}=1$) with large isolation (experimentally demonstrated 30 dB) in a broad intensity range, thus overcoming the trade-off mentioned above. This idea has been recently exploited for nonreciprocal double metasurfaces comprising one Lorentz and one Fano nonlinear dielectric metasurfaces[251].

An interesting realization of this nonlinear and asymmetric nonreciprocal Lorentz-Fano concept has been recently proposed for chip-based LiDAR technology[252]. Remarkably, the system architecture has been optimized by so-called inverse design, as currently used for designing complex electrodynamic systems[253]. It has been demonstrated that $\chi^{(3)}$ nonlinear resonators can be used to achieve fully passive, bias-free nonreciprocal pulse routing in standard silicon photonic platforms. To overcome the constraints of Eq.(14), this system consists of cascaded nonlinear Fano and Lorentzian resonators. As a result, the quasi-unitary forward transmission, along with a broad operating power range, has been reported, making this design very interesting for various applications, including LiDAR systems.

Thus, the nonlinear nonreciprocal structures are exciting and promising for practical applications in nonreciprocal devices over the entire spectrum. However, it worth noting that since this approach relies on the asymmetry of the field distribution over the structure, the simultaneous excitation from both ports will generally lead to a reduction of the effect. This fact suggests that *nonlinear isolators are unsuited for protecting continuous-wave sources from reflections*. Nonetheless, they can still be used to protect pulsed sources provided that the reflected pulse reaches the isolator during the off period of the source. For use in the continuous wave excitation scenario, one has to take care of the absence of the back-reflected waves, for example, using terminators or perfect absorbers.

We conclude this section with a feasible design of a nonlinear nonreciprocal dielectric structure. In 2011 Krasnok *et al.* proposed a new type of optical antenna based on dielectric particles with high dielectric constant and low losses[254,255]. It was shown that a single dielectric particle made of a high-index material can have the properties of a *Huygens element*[254]. In addition to electrical resonances, dielectric particles have strong magnetic Mie resonances that can be used to create Yagi-Uda nanoantennas for the optical wavelength range[254–258]. Later on, the concept of superdirective antennas was proposed based on the excitation of higher-order magnetic multipole moments in subwavelength dielectric particles, and the prototype was experimentally implemented[259]. In addition to the high directivity, these antennas demonstrate the ability to control radiation at the subwavelength scale, which occurs due to the subwavelength radiation beam's sensitivity to changes in the position of the source[259–261]. Also, discrete microwave waveguides with excellent waveguiding properties[262] and oligomer structures with the Fano-resonance[263] have been realized based on such high-index dielectric resonators. Such all-dielectric systems become a powerful platform for all-optical light modulation and reconfiguration in the optical range[264–266], holding a great promise for all-optical nonreciprocal devices.

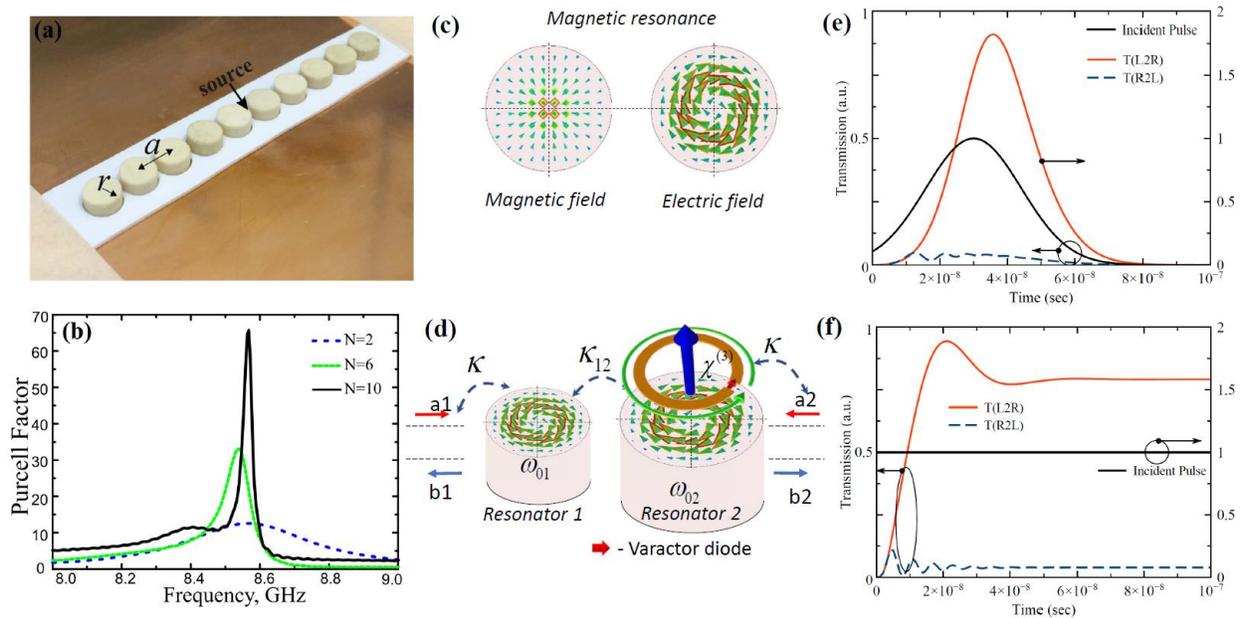

**Figure 7. Nonreciprocal system based on dielectric resonators with magnetic resonance.** (a) Antenna is composed of high-index dielectric subwavelength resonators (dielectric disks)[65]. (b) The proof-of-concept experiment at microwaves (~8.5 GHz) demonstrates 65-fold enhancement

of the Purcell factor in a chain of 10 dielectric particles. (c) Profile of electric and magnetic field at the magnetic dipole resonance. (d) Illustration of the feasible nonreciprocal structure that is under realization at *RadiaBeam* facility. (e,f) Typical results for (e) a Gaussian incoming pulse and (f) monochromatic wave.

Figure 7(a) exemplifies one of the microwave designs of an antenna composed of high-index dielectric subwavelength resonators (dielectric disks)[65]. In this work, Krasnok *et al*. have reported extraordinary enhancement of the Purcell effect governed by the so-called Van Hove singularity, Figure 7(b). Namely, the proof-of-concept experiment at ~8.5 GHz demonstrates a 65-fold enhancement of the Purcell factor in a chain of 10 dielectric particles.

The electric and magnetic field profile at the magnetic dipole resonance of a high-index disk dielectric resonator is shown in Figure 7(c). One can see that the electric field profile at this resonance is similar to metallic coils. Namely, it consists of two degenerate modes (clock- and counter-clockwise polarized circular field distribution). Breaking the modes' degeneracy would lead to the nonreciprocity effect, similar to devices based on split-ring resonators (see above).

In order to break the degeneracy, we suggest using a varactor diode connected to a split-ring resonator (SSR). The SSR is coupled with one of the dielectric discs. The resonance of the SSR is set to the magnetic resonance of the disk. The resulting system consists of a linear resonator#1 and nonlinear resonator#2, Figure 7(d). The coupling of resonator#2 to the nonlinear SSR provides the asymmetry required for the nonlinearity-based nonreciprocity. We have analyzed this system using the temporal coupled-mode theory (CMT) [see, for example, in Ref.[267]]. Following the CMT approach, we assume the system is excited by the left (right) channel with amplitudes of incoming and outgoing waves of $a_1$ and $b_1$ ($a_2$ and $b_2$), respectively. The coupling rates of both channels to the structure are assumed to be the same, $\kappa = 5/f$, where $f$ is the excitation frequency, equal to 1 GHz in our model. The eigenfrequencies of both resonators are also equal to 1 GHz. The intrinsic decay rates of the resonators are $f/40$, giving the Q-factors of 125, which is typical for lossless high-index microwave ceramics. The results for a Gaussian incoming pulse and monochromatic wave are presented in Figure 7(e,f). The wave transmission through the structure from left to right (L2R) is much larger than the opposite propagation (R2L), manifesting the nonreciprocal operation regime. Note that this single nonlinear resonator system suffers from the nonreciprocity-transmission tradeoff discussed above, Eq. (14). As was demonstrated in Ref.[250], this system can be significantly improved using a nonlinear Fano resonator coupled to a nonlinear Lorenzian resonator.

## 7. Nonreciprocity in quantum systems

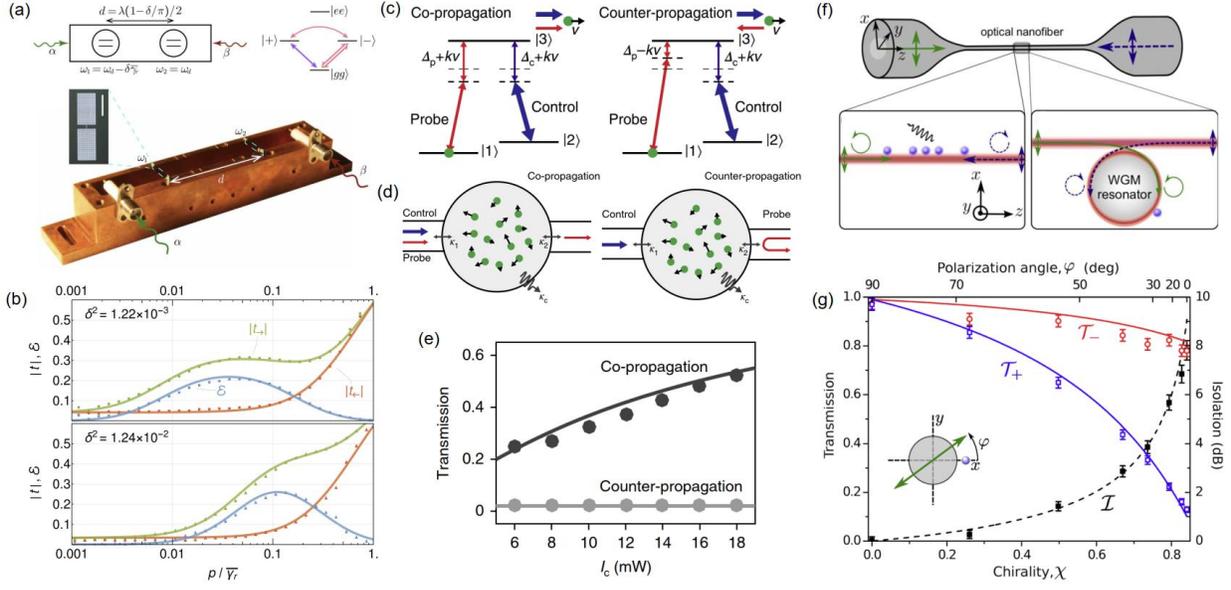

**Figure 8. Nonreciprocal quantum systems.** (a),(b) Nonreciprocity realized with quantum nonlinearity in superconductive Josephson junctions[65]. (a) (left) Schematic of the quantum diode: two qubits embedded in a 1D waveguide, tuned to the optimal conditions for nonreciprocal behavior. An incoming field from the forward direction (α drive) at frequency ωd is partially transmitted through the system, whereas a field incoming from the reverse direction (β drive) is fully reflected. (right) Energy level diagram of the system. (bottom) Open 1D waveguide with embedded 3D transmons (dashed mint green boxes). Inset: Optical micrograph of one of the two identical 3D transmons. (b) Nonreciprocity dependence on power for two detunings. Points are experimental data; solid lines are theoretical fits. (c) – (e) Thermal-motion-induced nonreciprocal quantum optical system[52]. (c),(d) In the presence of thermal motion, for the atom with a velocity component v along the laser beam: the atoms 'see' the same 'microscopic' Doppler shifts in the co-propagation case (left; high absorption) or the atoms 'see' opposite 'microscopic' Doppler shifts in the counter-propagation case (right; low absorption). (e) Normalized on-resonance transmission versus the control field power. The black (grey) dots are for the co-propagation (counter-propagation) case. (f),(g) Nanophotonic optical isolator controlled by the internal state of cold atoms[62]. (f) Optical nanofiber with trapped atoms either in the vicinity of the nanofiber or coupled to a whispering-gallery-mode (WGM). (g) Nonreciprocal transmission of chiral photons that interact with an ensemble of spin-polarized atoms. Inset: Cross section of the optical nanofiber (gray disk) including the trapped atoms (blue sphere) and the main polarization axis of the guided field (green double arrow).

Nonreciprocal systems in the quantum realm are of paramount interest in quantum computing and quantum networks[268]. The existing approaches to quantum nonreciprocity are presented in Figure 8. These are approaches based on quantum nonlinearity (saturation of two-level system), unidirectional loss or gain, and chiral coupling of light and matter.

Schemes for building nonreciprocal devices based on quantum nonlinearity have been proposed in Refs.[67,269,270] and attracted significant attention since then [66,271–273]. The quantum theory of such quantum nonreciprocal systems, also known as quantum diodes, has been presented in Refs.[271,272]. Subsequent works have unfolded the mechanism of this nonreciprocity, determined the analytical bounds for its efficiency, and revealed that entanglement between the atoms or qubits and the electromagnetic field is crucial for the nonreciprocal behavior[273]. This type of quantum nonreciprocity has been realized with two superconductive Josephson junctions in the microwave spectral range[65], Figure 8(a),(b). When the thermal noise can be neglected at cryogenic temperatures, energy states of a conventional LC contour get quantized, that is, gets the spectrum of a harmonic oscillator. The introduction of a lossless superconducting Josephson junction, operating as a nonlinear inductance, into such an LC contour makes the system effectively two-level as the other states are detuned from the main excitation[13,14]. A parallel connection of two Josephson junctions makes the resulting system tunable via external magnetic field bias, giving rise to the SQUID architecture.

In the work Ref.[65], Figure 8(a),(b), two such tunable SQUID qubits have been placed in a microwave waveguide that provided the desired phase retardation between the qubits. For a certain phase difference, the resulting system is described by two dressed states, a bright state and a purely dark state. The biasing of one qubit [first one in Figure 8(a)] provides the asymmetry desired for this sort of nonreciprocity. The quasi-dark state |+> can be populated by the driving field either directly from the ground state |gg> (purple path) or indirectly through the bright state |−> (pink path). These two channels interfere either constructively or destructively, depending on the driving direction. If interfering constructively, part of the population gets trapped in the quasi-dark state |+>, which in turn gives rise to the nonreciprocal behavior of the system. The system is predominantly trapped in the quasi-dark state |+> and is therefore partially transparent to the incident signal due to the extremely low saturability of |+>. Figure 8(b) demonstrates the analytical and experimental results for the leftward transmission, rightward transmission, and the isolation factor ($\varepsilon$) for two values of the detuning ($\delta$). For a review of microwave circulators at the quantum limit based on Josephson junctions, we refer the reader to Ref.[11].

Another type of quantum nonreciprocity is the one based on unidirectional gain or loss. For example, in Ref.[52] an interesting approach based on absorption asymmetry and electromagnetically induced transparency (EIT) in a gas of freely moving atoms has been proposed, Figure 8(c),(d). In the field of a control laser, the 3-level $\Lambda$ atoms are getting excited to the upper state reducing the absorption of the probe laser field (EIT effect). In the case of moving atoms, the control and probe field frequency get changed by the Doppler shift. The sight of this shift for each atom depends on whether the laser beams are co- or counter-propagating. As a result, the system operates in the EIT regime with high transmission for the co-propagating beams, whereas the counter-propagating results in strong absorption. The stronger intensity of the control beam, the larger population of the upper state and hence the weaker (stronger) absorption (transmission), Figure 8(e). Remarkably, this nonreciprocity effect gets increased with temperature growth and vanishes at low temperatures. A related approach, but utilizing the drift current biasing of graphene, has been theoretically proposed in Ref.[58].

Chiral coupling between waveguide modes and quantum emitters provides another perspective approach to quantum nonreciprocity[61,62,274–276]. Provoked by recent remarkable experimental achievements in the area of waveguide-atom (QDs) coupling[277–279], this approach is of great interest today. The essence of this approach is as follows. Evanescent fields of various waveguides can have a chiral property with circular polarization defined by the propagation direction. If a chiral atom (molecule, QD, qubit) with different oscillator strength of transitions to the states with m=-1, m=+1 is coupled to such a chiral evanescent field, it experiences different scattering efficiency depending on the mode propagation direction. As discussed in Refs.[276,280–282], this chiral interaction between the waveguiding mode and chiral atoms results in nonreciprocal transmission.

Figure 8(f) exemplifies this approach[62]. In this work, the chiral coupling of transverse spin in the evanescent field of guiding light with spin-polarized quantum emitters has been utilized for the magnet-less quantum nonreciprocal system. The larger the chirality, the stronger the isolation factor, which is shown to reach the value of 10 dB along with a quite high forward transmission of ~0.8, Figure 8(g). The coupling strength of the mode and atom is controlled by the polarization angle ($\varphi$). Remarkably, it has been shown that the spin of emitters can control the direction of the optical diode.

An approach that exploits both unidirectional gain effect and chiral light-matter coupling has been proposed in Ref.[53]. This work takes advantage of enhanced stimulated Raman scattering of crystalline Si[283,284] for Raman amplification of circularly polarized light. The Raman amplification is spin-locked by choice of the spin of the pumping field. As a result, if the spin of the signal beam coincides with the pump, the signal beam gets amplified. Otherwise, it gets absorbed, giving rise to the time-symmetry breaking and nonreciprocal effect.

Finally, in Ref.[193] a similar approach based on the valley polarization has been proposed. Today, valleytronics attracts a great deal of interest due to the discovery of atomically thin transition metal dichalcogenides (TMDCs)[285], whose broken inversion symmetry, in combination with the time-reversal symmetry, leads to opposite spins at the +K and -K valleys, effectively locking the spin and valley degrees of freedom[145,286–289]. As a result, if a TMDC is excited by, let us say, clockwise polarization, the excited excitons get this polarization and keep it for a while. This valley polarization acts as an effective magnetic field breaking the time-reversal symmetry[193]. This approach is relative to the one based on chiral coupling but does not require the chirality of field or any waveguide modes. Also, TMDC materials are more feasible for experimental work than atoms that typically require laser cooling.

## Conclusions

In this work, we have reviewed the up-and-coming advances in nonreciprocity. We have briefly discussed the physical picture that lies behind nonreciprocal materials and systems, providing the major characteristics that describe nonreciprocal systems. We have briefly discussed the magnet-based nonreciprocal systems, their disadvantages, and the current trends in this area, mainly associated with exploring and using novel materials, topological insulators, Weyl semimetals, metasurfaces, magneto-optical materials boosted plasmonic resonances, and 2D gyrotropic materials. Next, we have surveyed nonreciprocal systems based on active structures, PT-symmetry breaking, time-modulation, and nonlinearity. We have finalized our review by discussing various approaches in quantum systems, including quantum nonlinearity, unidirectional gain/loss, and chiral light-matter coupling. Throughout the work, we have discussed the pros and cons of the mentioned approaches to nonreciprocity. This analysis has allowed us to suggest a design of a feasible nonreciprocal structure that is currently under investigation at *RadiaBeam*. The design consists of two microwave resonators made of high-index microwave ceramics. The system is assumed to be asymmetric, and at least one of the resonators should be nonlinear. These two assumptions allow the resulting structure to be nonreciprocal. The proposed nonlinear and nonreciprocal structure's behavior has been analyzed in the temporal coupled-mode theory framework.

## Acknowledgment

This work was supported by the U.S. Department of Energy, Office of High Energy Physics, under contract DE-SC0020926. We would also like to thank Dr. Paul Carriere and Dr. Nikita Nefedkin for their valuable feedback.

## Authors information

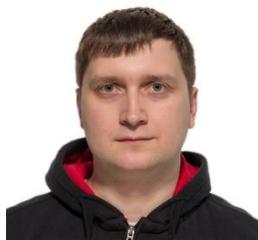

**Dr. Sergey V. Kutsaev** received his Ph.D. in accelerator physics from National Research Nuclear University "MEPhI", Moscow, Russia in 2011, under supervision of Prof. Nikolai Sobenin for his work on the accelerators for novel cargo inspection systems. Then, Dr. Kutsaev worked at Argonne National Laboratory, USA on heavy-ion and superconducting accelerators. Since 2014 he serves as an RF Group Leader at RadiaBeam Technologies, USA. His research interests include industrial accelerators, microwave and terahertz systems, RF superconducting and quantum technologies.

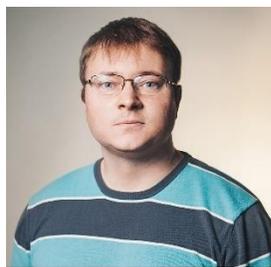

**Dr. Alex Krasnok** received his Ph.D. in optics from ITMO University, Russia. From 2013 to 2016, Dr. Krasnok has served as the head of the all-dielectric group at the ITMO University. From 2016 to 2018, Dr. Krasnok was a research scientist at the University of Texas at Austin. Since 2018, he is an Assistant Professor and core facility director at City University of New York (CUNY), where he has successfully established a new research program in photonics and created a photonics laboratory from scratch. His research interests span the fields of photonics, quantum optics, and metamaterials with a particular focus on interdisciplinarity and innovations. He has made significant contributions in the areas of scattering control, nanoantennas, low-loss dielectric nanostructures, metasurfaces, and optics of 2D transition-metal dichalcogenides.

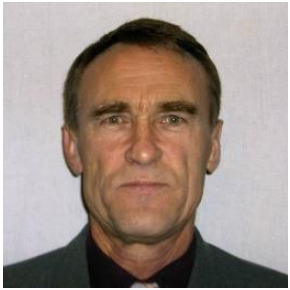
**Dr. Sergey N. Romanenko** received his Ph.D. in Dnepropetrovsk National University, Ukraine in 2000 in radiophysics. From 1981 to the present, Dr. Romanenko works at National University "Zaporizhzhia Polytechnic", Zaporizhzhia, Ukraine. His main scientific interests are electromagnetic field theory, complex media electromagnetics, antenna theory and microwave engineering.

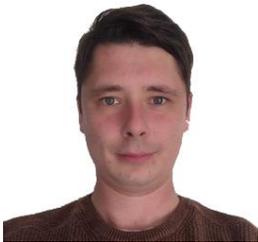
**Dr. Alexander Yu. Smirnov** received his Ph.D. in accelerator physics from National Research Nuclear University "MEPhI", Moscow, Russia in 2014, for his work on the RF deflector for XFEL, Germany. Dr. Smirnov worked for Siemens AG from 2011 to 2016 on high-power solid-state RF generators. Since 2016 he works at RadiaBeam Technologies, currently as an Accelerator Scientist. His research interests include RF amplifiers, particle accelerators, meta-materials and non-linear beam optics.

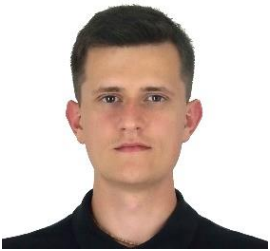
**Kirill Taletski** received his Master's in applied physics from National Research Nuclear University "MEPhI", Moscow, Russia in 2019, under the supervision of Prof. Maria Gusarova and Dr. Sergey Kutsaev for his work on the superconducting niobium RF cavities for quantum information systems. Since 2018 Mr. Taletski works at RadiaBeam Technologies, currently as a research associate. His research interests include microwave systems, RF superconductivity for accelerators and quantum technologies.

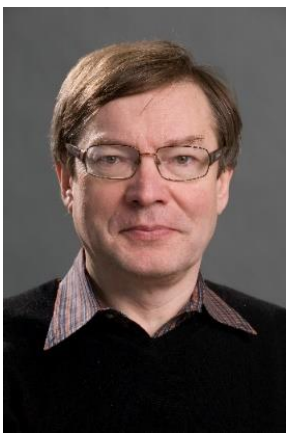
**Dr. Vyacheslav P. Yakovlev** received MS degree in accelerator physics from Novosibirsk State University (NSU), Russia, in 1977, and PhD in accelerator physics from Budker Institute for Nuclear Physics (Budker INP), Novosibirsk, Russia, in 1988. From 1994 to 1996, he was an Associate Professor at Novosibirsk State Technical University, and from 1988 to 1996, a Senior Scientist and a Group Leader at Budker INP. Since 1996 he has worked at Yale Beam Lab, Physics Department, Yale University, and Omega-P Inc. Currently, he is the Head of SRF Development Department at Application Science and Technology Division of Fermilab. From 2017 to present, Dr. Yakovlev serves as an Adjunct Professor of Accelerator Science, Facility for Rare Isotope Beams, Michigan State University, Lansing, USA. The scope of his professional interest includes physics and techniques of particle accelerators, namely: theory and simulations of the fields and beam dynamics in linear and circular accelerators; physics and technique of RF accelerator structures including room temperature cavities and stricture, superconducting cavities, and ferrite-tuned cavities; high power RF systems and RF sources for accelerators; tuning systems and cryo-module design.

# References

[1] I. A. Sukhoivanov, I. V. Guryev, *Photonic Crystals*, Springer Berlin Heidelberg, Berlin, Heidelberg, **2009**.
[2] J. D. Joannopoulos, S. G. Johnson, J. N. Winn, R. D. Meade, *Photonic Crystals: Molding the Flow of Light*, Princeton University Press, **2011**.
[3] Stefan Alexander Maier, *Plasmonics: Fundamentals and Applications*, Springer Science & Business Media, **2007**.
[4] A. I. Kuznetsov, A. E. Miroshnichenko, M. L. Brongersma, Y. S. Kivshar, B. Luk'yanchuk, B. Luk'yanchuk, *Science (80-. ).* **2016**, *354*, aag2472.
[5] A. Krasnok, M. Caldarola, N. Bonod, A. Alú, *Adv. Opt. Mater.* **2018**, *6*, 1701094.
[6] S. B. Glybovski, S. A. Tretyakov, P. A. Belov, Y. S. Kivshar, C. R. Simovski, *Phys. Rep.* **2016**, *634*, 1.
[7] F. Monticone, A. Alu, *Reports Prog. Phys.* **2017**, *80*, DOI 10.1088/1361-6633/aa518f.
[8] H. T. Chen, A. J. Taylor, N. Yu, *Reports Prog. Phys.* **2016**, *79*, 0.
[9] L. Lu, J. D. Joannopoulos, M. Soljačić, *Nat. Photonics* **2014**, *8*, 821.
[10] F. Arute, K. Arya, R. Babbush, D. Bacon, J. C. Bardin, R. Barends, R. Biswas, S. Boixo, F. G. S. L. Brandao, D. A. Buell, B. Burkett, Y. Chen, Z. Chen, B. Chiaro, R. Collins, W. Courtney, A. Dunsworth, E. Farhi, B. Foxen, A. Fowler, C. Gidney, M. Giustina, R. Graff, K. Guerin, S. Habegger, M. P. Harrigan, M. J. Hartmann, A. Ho, M. Hoffmann, T. Huang, T. S. Humble, S. V. Isakov, E. Jeffrey, Z. Jiang, D. Kafri, K. Kechedzhi, J. Kelly, P. V. Klimov, S. Knysh, A. Korotkov, F. Kostritsa, D. Landhuis, M. Lindmark, E. Lucero, D. Lyakh, S. Mandrà, J. R. McClean, M. McEwen, A. Megrant, X. Mi, K. Michielsen, M. Mohseni, J. Mutus, O. Naaman, M. Neeley, C. Neill, M. Y. Niu, E. Ostby, A. Petukhov, J. C. Platt, C. Quintana, E. G. Rieffel, P. Roushan, N. C. Rubin, D. Sank, K. J. Satzinger, V. Smelyanskiy, K. J. Sung, M. D. Trevithick, A. Vainsencher, B. Villalonga, T. White, Z. J. Yao, P. Yeh, A. Zalcman, H. Neven, J. M. Martinis, *Nature* **2019**, *574*, 505.
[11] L. Ranzani, J. Aumentado, *IEEE Microw. Mag.* **2019**, *20*, 112.
[12] W. Bogaerts, D. Pérez, J. Capmany, D. A. B. Miller, J. Poon, D. Englund, F. Morichetti, A. Melloni, *Nature* **2020**, *586*, 207.
[13] X. Gu, A. F. Kockum, A. Miranowicz, Y. Liu, F. Nori, *Phys. Rep.* **2017**, *718–719*, 1.
[14] A. Blais, S. M. Girvin, W. D. Oliver, *Nat. Phys.* **2020**, *16*, 247.
[15] S. V. Kutsaev, K. Taletski, R. Agustsson, P. Carriere, A. N. Cleland, Z. A. Conway, É. Dumur, A. Moro, A. Y. Smirnov, *EPJ Quantum Technol.* **2020**, *7*, 7.
[16] C. Caloz, A. Alù, S. Tretyakov, D. Sounas, K. Achouri, Z.-L. Deck-Léger, *Phys. Rev. Appl.* **2018**, *10*, 047001.
[17] C. Caloz, Z.-L. Deck-Leger, *IEEE Trans. Antennas Propag.* **2020**, *68*, 1569.

[18]  C. Caloz, Z.-L. Deck-Leger, *IEEE Trans. Antennas Propag.* **2019**, *68*, 1.
[19]  D. L. Sounas, A. Alù, *Nat. Photonics* **2017**, *11*, 774.
[20]  S. Fan, Y. Shi, Q. Lin, *IEEE Antennas Wirel. Propag. Lett.* **2018**, *17*, 1948.
[21]  N. A. Estep, D. L. Sounas, J. Soric, A. Alù, *Nat. Phys.* **2014**, *10*, 923.
[22]  H. Lira, Z. Yu, S. Fan, M. Lipson, *Phys. Rev. Lett.* **2012**, *109*, 033901.
[23]  K. Fang, Z. Yu, S. Fan, *Phys. Rev. Lett.* **2012**, *108*, 153901.
[24]  D. L. Sounas, A. Alù, *Nat. Photonics* **2017**, *11*, 774.
[25]  K. Fang, Z. Yu, S. Fan, *Nat. Photonics* **2012**, *6*, 782.
[26]  K. Fang, J. Luo, A. Metelmann, M. H. Matheny, F. Marquardt, A. A. Clerk, O. Painter, *Nat. Phys.* **2017**, *13*, 465.
[27]  L. D. Tzuang, K. Fang, P. Nussenzveig, S. Fan, M. Lipson, *Nat. Photonics* **2014**, *8*, 701.
[28]  R. Fleury, D. L. Sounas, C. F. Sieck, M. R. Haberman, A. Alu, *Science (80-. ).* **2014**, *343*, 516.
[29]  D.-W. Wang, H.-T. Zhou, M.-J. Guo, J.-X. Zhang, J. Evers, S.-Y. Zhu, *Phys. Rev. Lett.* **2013**, *110*, 093901.
[30]  L. Fan, J. Wang, L. T. Varghese, H. Shen, B. Niu, Y. Xuan, A. M. Weiner, M. Qi, *Science (80-. ).* **2012**, *335*, 447.
[31]  X. Guo, C.-L. Zou, H. Jung, H. X. Tang, *Phys. Rev. Lett.* **2016**, *117*, 123902.
[32]  M. Lawrence, D. R. Barton, J. A. Dionne, *Nano Lett.* **2018**, *18*, 1104.
[33]  A. B. Khanikaev, A. Alù, *Nat. Photonics* **2015**, *9*, 359.
[34]  Z. Yu, S. Fan, *Appl. Phys. Lett.* **2009**, *94*, 171116.
[35]  Z. Yu, S. Fan, *Nat. Photonics* **2009**, *3*, 91.
[36]  F. Ruesink, M. A. Miri, A. Alù, E. Verhagen, *Nat. Commun.* **2016**, *7*, 1.
[37]  M. Hafezi, P. Rabl, *Opt. Express* **2012**, *20*, 7672.
[38]  Z. Shen, Y.-L. Zhang, Y. Chen, C.-L. Zou, Y.-F. Xiao, X.-B. Zou, F.-W. Sun, G.-C. Guo, C.-H. Dong, *Nat. Photonics* **2016**, *10*, 657.
[39]  N. R. Bernier, L. D. Tóth, A. Koottandavida, M. A. Ioannou, D. Malz, A. Nunnenkamp, A. K. Feofanov, T. J. Kippenberg, *Nat. Commun.* **2017**, *8*, 604.
[40]  F. Ruesink, J. P. Mathew, M.-A. Miri, A. Alù, E. Verhagen, *Nat. Commun.* **2018**, *9*, 1798.
[41]  M. S. Kang, A. Butsch, P. S. J. Russell, *Nat. Photonics* **2011**, *5*, 549.
[42]  D. B. Sohn, S. Kim, G. Bahl, *Nat. Photonics* **2018**, *12*, 91.
[43]  L. Chang, X. Jiang, S. Hua, C. Yang, J. Wen, L. Jiang, G. Li, G. Wang, M. Xiao, *Nat. Photonics* **2014**, *8*, 524.
[44]  N. Bender, S. Factor, J. D. Bodyfelt, H. Ramezani, D. N. Christodoulides, F. M. Ellis, T. Kottos, *Phys. Rev. Lett.* **2013**, *110*, 234101.
[45]  B. Peng, S. K. Özdemir, F. Lei, F. Monifi, M. Gianfreda, G. L. Long, S. Fan, F. Nori, C. M. Bender, L. Yang, Ş. K. Özdemir, F. Lei, F. Monifi, M. Gianfreda, G. L. Long, S. Fan, F. Nori, C. M. Bender, L. Yang, *Nat. Phys.* **2014**, *10*, 394.
[46]  S. Hua, J. Wen, X. Jiang, Q. Hua, L. Jiang, M. Xiao, *Nat. Commun.* **2016**, *7*, 13657.
[47]  Y. Zheng, J. Yang, Z. Shen, J. Cao, X. Chen, X. Liang, W. Wan, *Light Sci. Appl.* **2016**, *5*, e16072.
[48]  B. He, L. Yang, X. Jiang, M. Xiao, *Phys. Rev. Lett.* **2018**, *120*, 203904.
[49]  C. Liang, B. Liu, A.-N. Xu, X. Wen, C. Lu, K. Xia, M. K. Tey, Y.-C. Liu, L. You, *Phys. Rev. Lett.* **2020**, *125*, 123901.
[50]  G. Lin, S. Zhang, Y. Hu, Y. Niu, J. Gong, S. Gong, *Phys. Rev. Lett.* **2019**, *123*, 033902.
[51]  B. Jin, C. Argyropoulos, *Adv. Opt. Mater.* **2019**, *7*, 1901083.

[52] S. Zhang, Y. Hu, G. Lin, Y. Niu, K. Xia, J. Gong, S. Gong, *Nat. Photonics* **2018**, *12*, 744.
[53] M. Lawrence, J. A. Dionne, *Nat. Commun.* **2019**, *10*, 3297.
[54] S. Maayani, R. Dahan, Y. Kligerman, E. Moses, A. U. Hassan, H. Jing, F. Nori, D. N. Christodoulides, T. Carmon, *Nature* **2018**, *558*, 569.
[55] R. Huang, A. Miranowicz, J. Q. Liao, F. Nori, H. Jing, *Phys. Rev. Lett.* **2018**, *121*, 153601.
[56] H. Lü, Y. Jiang, Y.-Z. Wang, H. Jing, *Photonics Res.* **2017**, *5*, 367.
[57] E. Sánchez-Burillo, A. González-Tudela, C. Gonzalez-Ballestero, *Phys. Rev. A* **2020**, *102*, 1.
[58] T. A. Morgado, M. G. Silveirinha, *ACS Photonics* **2018**, *5*, 4253.
[59] L. Tang, J. Tang, W. Zhang, G. Lu, H. Zhang, Y. Zhang, K. Xia, M. Xiao, *Phys. Rev. A* **2019**, *99*, 043833.
[60] W. Bin Yan, W. Y. Ni, J. Zhang, F. Y. Zhang, H. Fan, *Phys. Rev. A* **2018**, *98*, 43852.
[61] C. Gonzalez-Ballestero, E. Moreno, F. J. Garcia-Vidal, A. Gonzalez-Tudela, *Phys. Rev. A* **2016**, *94*, 063817.
[62] C. Sayrin, C. Junge, R. Mitsch, B. Albrecht, D. O'Shea, P. Schneeweiss, J. Volz, A. Rauschenbeutel, *Phys. Rev. X* **2015**, *5*, 041036.
[63] N. Almeida, T. Werlang, D. Valente, *J. Opt. Soc. Am. B* **2019**, *36*, 3357.
[64] P. Yang, X. Xia, H. He, S. Li, X. Han, P. Zhang, G. Li, P. Zhang, J. Xu, Y. Yang, T. Zhang, *Phys. Rev. Lett.* **2019**, *123*, 233604.
[65] A. Rosario Hamann, C. Müller, M. Jerger, M. Zanner, J. Combes, M. Pletyukhov, M. Weides, T. M. Stace, A. Fedorov, *Phys. Rev. Lett.* **2018**, *121*, 123601.
[66] E. Mascarenhas, M. F. Santos, A. Auffèves, D. Gerace, *Phys. Rev. A* **2016**, *93*, 043821.
[67] D. Roy, *Phys. Rev. B* **2010**, *81*, 155117.
[68] K. M. Sliwa, M. Hatridge, A. Narla, S. Shankar, L. Frunzio, R. J. Schoelkopf, M. H. Devoret, *Phys. Rev. X* **2015**, *5*, 041020.
[69] F. Lecocq, L. Ranzani, G. A. Peterson, K. Cicak, R. W. Simmonds, J. D. Teufel, J. Aumentado, *Phys. Rev. Appl.* **2017**, *7*, 024028.
[70] C. Müller, S. Guan, N. Vogt, J. H. Cole, T. M. Stace, *Phys. Rev. Lett.* **2018**, *120*, 213602.
[71] J. Kerckhoff, K. Lalumière, B. J. Chapman, A. Blais, K. W. Lehnert, *Phys. Rev. Appl.* **2015**, *4*, 034002.
[72] G. Viola, D. P. DiVincenzo, *Phys. Rev. X* **2014**, *4*, 021019.
[73] H. Nassar, B. Yousefzadeh, R. Fleury, M. Ruzzene, A. Alù, C. Daraio, A. N. Norris, G. Huang, M. R. Haberman, *Nat. Rev. Mater.* **2020**, *5*, 667.
[74] M. Brandenbourger, X. Locsin, E. Lerner, C. Coulais, *Nat. Commun.* **2019**, *10*, 4608.
[75] T. Van Mechelen, Z. Jacob, *Nanophotonics* **2019**, *8*, 1399.
[76] C. Caloz, A. Alù, S. Tretyakov, D. Sounas, K. Achouri, Z.-L. Deck-Léger, *Phys. Rev. Appl.* **2018**, *10*, 047001.
[77] A. B. Khanikaev, A. Alù, *Nat. Photonics* **2015**, *9*, 359.
[78] Y. Hadad, J. C. Soric, A. Alu, *Proc. Natl. Acad. Sci.* **2016**, *113*, 3471.
[79] R. J. Potton, *Reports Prog. Phys.* **2004**, *67*, 717.
[80] D. L. Sounas, C. Caloz, A. Alù, *Nat. Commun.* **2013**, *4*, 2407.
[81] A. Kord, D. L. Sounas, A. Alu, *Proc. IEEE* **2020**, *108*, 1728.
[82] D. Jalas, A. Petrov, M. Eich, W. Freude, S. Fan, Z. Yu, R. Baets, M. Popović, A. Melloni, J. D. Joannopoulos, M. Vanwolleghem, C. R. Doerr, H. Renner, *Nat. Photonics* **2013**, *7*, 579.

[83] D. G. Baranov, Y. Xiao, I. A. Nechepurenko, A. Krasnok, A. Alù, M. A. Kats, *Nat. Mater.* **2019**, *18*, 920.
[84] A. Kord, D. L. Sounas, A. Alu, *IEEE Trans. Microw. Theory Tech.* **2018**, *66*, 911.
[85] A. Krasnok, D. Baranov, H. Li, M.-A. Miri, F. Monticone, A. Alú, *Adv. Opt. Photonics* **2019**, *11*, 892.
[86] M. J. Freiser, *IEEE Trans. Magn.* **1968**, *MAG-4*, 152.
[87] L. D. Landau, E. M. Lifshitz, L. P. Pitaevskii, *Electrodynamics of Continous Media*, Pergamon, New York, **1984**.
[88] G. Barzilai, G. Gerosa, *Proc. Inst. Electr. Eng.* **1966**, *113*, 285.
[89] P. J. Allen, *IEEE Trans. Microw. Theory Tech.* **1956**, *4*, 223.
[90] M. G. Silveirinha, *Opt. Express* **2019**, *27*, 14328.
[91] A. G. Fox, S. E. Miller, M. T. Weiss, *Bell Syst. Tech. J.* **1955**, *34*, 5.
[92] S. I. Shams, M. Elsaadany, A. A. Kishk, *IEEE Trans. Microw. Theory Tech.* **2019**, *67*, 94.
[93] S. Chen, F. Fan, X. Wang, P. Wu, H. Zhang, S. Chang, *Opt. Express* **2015**, *23*, 1015.
[94] S. Chen, F. Fan, X. He, M. Chen, S. Chang, *Appl. Opt.* **2015**, *54*, 9177.
[95] R. Kononchuk, C. Pfeiffer, I. Anisimov, N. I. Limberopoulos, I. Vitebskiy, A. A. Chabanov, *Phys. Rev. Appl.* **2019**, *12*, 054046.
[96] L. J. Aplet, J. W. Carson, *Appl. Opt.* **1964**, *3*, 544.
[97] S. Fischer, *J. Opt. Commun.* **1987**, *8*, DOI 10.1515/JOC.1987.8.1.18.
[98] J. Y. Chin, T. Steinle, T. Wehlus, D. Dregely, T. Weiss, V. I. Belotelov, B. Stritzker, H. Giessen, *Nat. Commun.* **2013**, *4*, 1599.
[99] V. V. Temnov, G. Armelles, U. Woggon, D. Guzatov, A. Cebollada, A. Garcia-Martin, J.-M. Garcia-Martin, T. Thomay, A. Leitenstorfer, R. Bratschitsch, *Nat. Photonics* **2010**, *4*, 107.
[100] F. Spitzer, A. N. Poddubny, I. A. Akimov, V. F. Sapega, L. Klompmaker, L. E. Kreilkamp, L. V. Litvin, R. Jede, G. Karczewski, M. Wiater, T. Wojtowicz, D. R. Yakovlev, M. Bayer, *Nat. Phys.* **2018**, *14*, 1043.
[101] E. G. Víllora, P. Molina, M. Nakamura, K. Shimamura, T. Hatanaka, A. Funaki, K. Naoe, *Appl. Phys. Lett.* **2011**, *99*, 011111.
[102] Z. Chen, L. Yang, Y. Hang, X. Wang, *Opt. Mater. (Amst).* **2015**, *47*, 39.
[103] J. G. Bai, G.-Q. Lu, T. Lin, *Sensors Actuators A Phys.* **2003**, *109*, 9.
[104] L. Weller, K. S. Kleinbach, M. A. Zentile, S. Knappe, I. G. Hughes, C. S. Adams, *Opt. Lett.* **2012**, *37*, 3405.
[105] W. Zhao, *Sensors Actuators A Phys.* **2001**, *89*, 250.
[106] D. Vojna, O. Slezák, A. Lucianetti, T. Mocek, *Appl. Sci.* **2019**, *9*, 3160.
[107] C. E. Fay, R. L. Comstock, *IEEE Trans. Microw. Theory Tech.* **1965**, *13*, 15.
[108] M. G. Barsukova, A. I. Musorin, A. S. Shorokhov, A. A. Fedyanin, *APL Photonics* **2019**, *4*, 016102.
[109] D. Zhong, K. L. Seyler, X. Linpeng, R. Cheng, N. Sivadas, B. Huang, E. Schmidgall, T. Taniguchi, K. Watanabe, M. A. McGuire, W. Yao, D. Xiao, K.-M. C. Fu, X. Xu, *Sci. Adv.* **2017**, *3*, e1603113.
[110] Y. Tokura, K. Yasuda, A. Tsukazaki, *Nat. Rev. Phys.* **2019**, *1*, 126.
[111] A. C. Mahoney, J. I. Colless, S. J. Pauka, J. M. Hornibrook, J. D. Watson, G. C. Gardner, M. J. Manfra, A. C. Doherty, D. J. Reilly, *Phys. Rev. X* **2017**, *7*, 011007.
[112] I. S. Maksymov, *Rev. Phys.* **2016**, *1*, 36.
[113] B. Sepúlveda, J. B. González-Díaz, A. García-Martín, L. M. Lechuga, G. Armelles, *Phys.*

*Rev. Lett.* **2010**, *104*, 147401.

[114] N. Maccaferri, I. Zubritskaya, I. Razdolski, I.-A. Chioar, V. Belotelov, V. Kapaklis, P. M. Oppeneer, A. Dmitriev, *J. Appl. Phys.* **2020**, *127*, 080903.

[115] Y. Li, Q. Zhang, A. V. Nurmikko, S. Sun, *Nano Lett.* **2005**, *5*, 1689.

[116] H. Uchida, Y. Masuda, R. Fujikawa, A. V. Baryshev, M. Inoue, *J. Magn. Magn. Mater.* **2009**, *321*, 843.

[117] A. V. Chetvertukhin, A. I. Musorin, T. V. Dolgova, H. Uchida, M. Inoue, A. A. Fedyanin, *J. Magn. Magn. Mater.* **2015**, *383*, 110.

[118] I. Zubritskaya, N. Maccaferri, X. Inchausti Ezeiza, P. Vavassori, A. Dmitriev, *Nano Lett.* **2018**, *18*, 302.

[119] V. Bonanni, S. Bonetti, T. Pakizeh, Z. Pirzadeh, J. Chen, J. Nogués, P. Vavassori, R. Hillenbrand, J. Åkerman, A. Dmitriev, *Nano Lett.* **2011**, *11*, 5333.

[120] V. I. Belotelov, I. A. Akimov, M. Pohl, V. A. Kotov, S. Kasture, A. S. Vengurlekar, A. V. Gopal, D. R. Yakovlev, A. K. Zvezdin, M. Bayer, *Nat. Nanotechnol.* **2011**, *6*, 370.

[121] A. Y. Frolov, M. R. Shcherbakov, A. A. Fedyanin, *Phys. Rev. B* **2020**, *101*, 045409.

[122] M. G. Barsukova, A. S. Shorokhov, A. I. Musorin, D. N. Neshev, Y. S. Kivshar, A. A. Fedyanin, *ACS Photonics* **2017**, *4*, 2390.

[123] G. P. Zouros, G. D. Kolezas, E. Almpanis, K. Baskourelos, T. P. Stefański, K. L. Tsakmakidis, *Nanophotonics* **2020**, *9*, 4033.

[124] Z. Sakotic, A. Krasnok, N. Cselyuszka, N. Jankovic, A. Alú, *Phys. Rev. Appl.* **2020**, *13*, DOI 10.1103/PHYSREVAPPLIED.13.064073.

[125] A. Krasnok, D. Baranov, H. Li, M.-A. Miri, F. Monticone, A. Alú, *Adv. Opt. Photonics* **2019**, *11*, 892.

[126] Y. Ra'di, A. Krasnok, A. Alù, *ACS Photonics* **2020**, *7*, 1468.

[127] F. Monticone, D. Sounas, A. Krasnok, A. Alù, *ACS Photonics* **2019**, *6*, acsphotonics.9b01104.

[128] V. A. Zakharov, A. N. Poddubny, *Phys. Rev. A* **2020**, *101*, 043848.

[129] A. A. Voronov, D. Karki, D. O. Ignatyeva, M. A. Kozhaev, M. Levy, V. I. Belotelov, *Opt. Express* **2020**, *28*, 17988.

[130] V. S. Asadchy, C. Guo, B. Zhao, S. Fan, *Adv. Opt. Mater.* **2020**, *8*, 2000100.

[131] B. Zhao, C. Guo, C. A. C. Garcia, P. Narang, S. Fan, *Nano Lett.* **2020**, *20*, 1923.

[132] N. P. Armitage, E. J. Mele, A. Vishwanath, *Rev. Mod. Phys.* **2018**, *90*, 015001.

[133] O. V. Kotov, Y. E. Lozovik, *Phys. Rev. B* **2018**, *98*, 195446.

[134] A. M. Shaltout, V. M. Shalaev, M. L. Brongersma, *Science (80-. ).* **2019**, *364*, DOI 10.1126/science.aat3100.

[135] M. Gibertini, M. Koperski, A. F. Morpurgo, K. S. Novoselov, *Nat. Nanotechnol.* **2019**, *14*, 408.

[136] T. Ozawa, H. M. Price, A. Amo, N. Goldman, M. Hafezi, L. Lu, M. C. Rechtsman, D. Schuster, J. Simon, O. Zilberberg, I. Carusotto, *Rev. Mod. Phys.* **2019**, *91*, 015006.

[137] Y. Tokura, N. Nagaosa, *Nat. Commun.* **2018**, *9*, DOI 10.1038/s41467-018-05759-4.

[138] M. Z. Hasan, C. L. Kane, *Rev. Mod. Phys.* **2010**, *82*, 3045.

[139] M. Z. Hasan, J. E. Moore, *Annu. Rev. Condens. Matter Phys.* **2011**, *2*, 55.

[140] D. Hsieh, D. Qian, L. Wray, Y. Xia, Y. S. Hor, R. J. Cava, M. Z. Hasan, *Nature* **2008**, *452*, 970.

[141] J. E. Moore, *Nature* **2010**, *464*, 194.

[142] C. L. Kane, E. J. Mele, *Phys. Rev. Lett.* **2005**, *95*, 146802.

[143] A. B. Khanikaev, S. Hossein Mousavi, W. K. Tse, M. Kargarian, A. H. MacDonald, G. Shvets, *Nat. Mater.* **n.d.**, *12*, 233.
[144] S. M. Young, S. Zaheer, J. C. Y. Teo, C. L. Kane, E. J. Mele, A. M. Rappe, *Phys. Rev. Lett.* **2012**, *108*, 1.
[145] L. Sun, C.-Y. Wang, A. Krasnok, J. Choi, J. Shi, J. S. Gomez-Diaz, A. Zepeda, S. Gwo, C.-K. Shih, A. Alù, X. Li, *Nat. Photonics* **2019**, *13*, 180.
[146] A. Chanana, N. Lotfizadeh, H. O. Condori Quispe, P. Gopalan, J. R. Winger, S. Blair, A. Nahata, V. V Deshpande, M. A. Scarpulla, B. Sensale-Rodriguez, *ACS Nano* **2019**, *13*, 4091.
[147] J. Hofmann, S. Das Sarma, **2016**, *241402*, 1.
[148] Z. Long, Y. Wang, M. Erukhimova, M. Tokman, A. Belyanin, *Phys. Rev. Lett.* **2018**, *120*, 37403.
[149] L. Lu, H. Gao, Z. Wang, *Nat. Commun.* **2018**, *9*, 1.
[150] Z. Sakotic, A. Krasnok, N. Cselyuszka, N. Jankovic, A. Alú, *Phys. Rev. Appl.* **2020**, *13*, 064073.
[151] L. Zhu, S. Fan, *Phys. Rev. B* **2014**, *90*, 220301.
[152] B. Zhao, Y. Shi, J. Wang, Z. Zhao, N. Zhao, S. Fan, *Opt. Lett.* **2019**, *44*, 4203.
[153] N. Yu, F. Capasso, *Nat. Mater.* **2014**, *13*, 139.
[154] C. L. Holloway, E. F. Kuester, J. A. Gordon, J. O'Hara, J. Booth, D. R. Smith, *IEEE Antennas Propag. Mag.* **2012**, *54*, 10.
[155] S. B. Glybovski, S. A. Tretyakov, P. A. Belov, Y. S. Kivshar, C. R. Simovski, *Phys. Rep.* **2016**, *634*, 1.
[156] S. Chang, X. Guo, X. Ni, *Annu. Rev. Mater. Res.* **2018**, *48*, 279.
[157] P. Genevet, F. Capasso, F. Aieta, M. Khorasaninejad, R. Devlin, *Optica* **2017**, *4*, 139.
[158] K. Koshelev, G. Favraud, A. Bogdanov, Y. Kivshar, A. Fratalocchi, *Nanophotonics* **2019**, *8*, 725.
[159] A. K. Iyer, A. Alu, A. Epstein, *IEEE Trans. Antennas Propag.* **2020**, *68*, 1223.
[160] A. Alù, *Nat. Mater.* **2016**, *15*, 1229.
[161] V. M. Shalaev, *Nat. Photonics* **2007**, *1*, 41.
[162] N. I. Zheludev, Y. S. Kivshar, *Nat. Mater.* **2012**, *11*, 917.
[163] N. Meinzer, W. L. Barnes, I. R. Hooper, *Nat. Photonics* **2014**, *8*, 889.
[164] D. McGrath, *IEEE Trans. Antennas Propag.* **1986**, *34*, 46.
[165] J. B. Pendry, **1996**, 4773.
[166] D. R. Smith, S. Schultz, P. Markoš, C. M. Soukoulis, *Phys. Rev. B - Condens. Matter Mater. Phys.* **2002**, *65*, 1.
[167] A. V. Kildishev, A. Boltasseva, V. M. Shalaev, *Science (80-. ).* **2013**, *339*, 1232009.
[168] P. Moitra, Y. Yang, Z. Anderson, I. I. Kravchenko, D. P. Briggs, J. Valentine, *Nat. Photonics* **2013**, *7*, 791.
[169] A. Krasnok, S. Makarov, M. Petrov, R. Savelev, P. Belov, **2015**, *9502*, 1.
[170] Y. Yang, I. I. Kravchenko, D. P. Briggs, J. Valentine, **2014**, *4826*, 1.
[171] K. E. Chong, B. Hopkins, I. Staude, A. E. Miroshnichenko, J. Dominguez, M. Decker, D. N. Neshev, I. Brener, Y. S. Kivshar, *Small* **2014**, *10*, 1985.
[172] A. I. Kuznetsov, A. E. Miroshnichenko, Y. H. Fu, J. Zhang, B. Luk'yanchuk, *Sci. Rep.* **2012**, *2*, 492.
[173] M. Khorasaninejad, F. Capasso, *Science (80-. ).* **2017**, *358*, 1.
[174] J. Engelberg, U. Levy, *Nat. Commun.* **2020**, *11*, 9.

[175] F. Ding, Y. Yang, R. A. Deshpande, S. I. Bozhevolnyi, *Nanophotonics* **2018**, *7*, 1129.
[176] X. Zang, H. Ding, Y. Intaravanne, L. Chen, Y. Peng, J. Xie, Q. Ke, A. V Balakin, A. P. Shkurinov, X. Chen, Y. Zhu, S. Zhuang, *Laser Photon. Rev.* **2019**, *13*, 1900182.
[177] S. Ali, H. Gangaraj, F. Monticone, S. A. H. Gangaraj, F. Monticone, *Nanophotonics* **2018**, *7*, 1025.
[178] A. Krasnok, M. Tymchenko, A. Alù, *Mater. Today* **2018**, *21*, 8.
[179] A. Krasnok, A. Alú, *J. Opt.* **2018**, *20*, 064002.
[180] J. M. Abendroth, M. L. Solomon, D. R. Barton, M. S. El Hadri, E. E. Fullerton, J. A. Dionne, *Adv. Opt. Mater.* **2020**, 2001420.
[181] A. I. Musorin, A. V. Chetvertukhin, T. V. Dolgova, H. Uchida, M. Inoue, B. S. Luk'yanchuk, A. A. Fedyanin, *Appl. Phys. Lett.* **2019**, *115*, 151102.
[182] E. Atmatzakis, N. Papasimakis, V. Fedotov, G. Vienne, N. I. Zheludev, *Nanophotonics* **2018**, *7*, 199.
[183] S. Taravati, B. A. Khan, S. Gupta, K. Achouri, S. Member, C. Caloz, *IEEE Trans. Antennas Propag.* **2017**, *65*, 3589.
[184] T. Kodera, D. L. Sounas, C. Caloz, *IEEE Antennas Wirel. Propag. Lett.* **2012**, *11*, 1454.
[185] Y. Ra'di, A. Grbic, *Phys. Rev. B* **2016**, *94*, 195432.
[186] A. M. Mahmoud, A. R. Davoyan, N. Engheta, *Nat. Commun.* **2015**, *6*, 8359.
[187] A. Shaltout, A. Kildishev, V. Shalaev, *Opt. Mater. Express* **2015**, *5*, 2459.
[188] H. Li, S. Ruan, Y. Zeng, *Adv. Mater.* **2019**, *31*, 1900065.
[189] B. Huang, G. Clark, E. Navarro-Moratalla, D. R. Klein, R. Cheng, K. L. Seyler, D. Zhong, E. Schmidgall, M. A. McGuire, D. H. Cobden, W. Yao, D. Xiao, P. Jarillo-Herrero, X. Xu, *Nature* **2017**, *546*, 270.
[190] Y. Deng, Y. Yu, Y. Song, J. Zhang, N. Z. Wang, Z. Sun, Y. Yi, Y. Z. Wu, S. Wu, J. Zhu, J. Wang, X. H. Chen, Y. Zhang, *Nature* **2018**, *563*, 94.
[191] C. Gong, L. Li, Z. Li, H. Ji, A. Stern, Y. Xia, T. Cao, W. Bao, C. Wang, Y. Wang, Z. Q. Qiu, R. J. Cava, S. G. Louie, J. Xia, X. Zhang, *Nature* **2017**, *546*, 265.
[192] T. Lan, B. Ding, B. Liu, *Nano Sel.* **2020**, *1*, 298.
[193] Y. Kawaguchi, S. Guddala, K. Chen, A. Alù, V. Menon, A. B. Khanikaev, *arXiv* **2020**, *arXiv:2007*, 1.
[194] I. Crassee, J. Levallois, A. L. Walter, M. Ostler, A. Bostwick, E. Rotenberg, T. Seyller, D. van der Marel, A. B. Kuzmenko, *Nat. Phys.* **2011**, *7*, 48.
[195] R. Shimano, G. Yumoto, J. Y. Yoo, R. Matsunaga, S. Tanabe, H. Hibino, T. Morimoto, H. Aoki, *Nat. Commun.* **2013**, *4*, 1841.
[196] J.-M. Poumirol, P. Q. Liu, T. M. Slipchenko, A. Y. Nikitin, L. Martin-Moreno, J. Faist, A. B. Kuzmenko, *Nat. Commun.* **2017**, *8*, 14626.
[197] D. Liu, S. Zhang, N. Ma, X. Li, *J. Appl. Phys.* **2012**, *112*, 023115.
[198] I. Crassee, M. Orlita, M. Potemski, A. L. Walter, M. Ostler, T. Seyller, I. Gaponenko, J. Chen, A. B. Kuzmenko, *Nano Lett.* **2012**, *12*, 2470.
[199] Y. Zhang, Y.-W. Tan, H. L. Stormer, P. Kim, *Nature* **2005**, *438*, 201.
[200] J. Maciejko, T. L. Hughes, S.-C. Zhang, *Annu. Rev. Condens. Matter Phys.* **2011**, *2*, 31.
[201] B. Placke, S. Bosco, D. DiVincenzo, *EPJ Quantum Technol.* **2017**, *4*, 5.
[202] D. M. Pozar, *Microwave Engineering, 4th Edition*, John Wiley & Sons, Inc., **2011**.
[203] J. G. Linvill, J. R. L. Wallace, *Negative Impedance Converters Employing Transistors*, **1955**, 2 726 370.
[204] T. Harrison, *IEEE Trans. Circuit Theory* **1963**, *10*, 303.

[205] A. Morse, L. Huelsman, *IEEE Trans. Circuit Theory* **1964**, *11*, 277.
[206] S. Tanaka, N. Shimomura, K. Ohtake, *Proc. IEEE* **1965**, *53*, 260.
[207] Y. Ayasli, *IEEE Trans. Magn.* **1989**, *25*, 3242.
[208] T. Kodera, D. L. Sounas, C. Caloz, *IEEE Antennas Wirel. Propag. Lett.* **2011**, *10*, 1551.
[209] Z. Wang, Z. Wang, J. Wang, B. Zhang, J. Huangfu, J. D. Joannopoulos, M. Soljacic, L. Ran, *Proc. Natl. Acad. Sci.* **2012**, *109*, 13194.
[210] S. W. Y. Mung, W. S. Chan, *IEEE Microw. Wirel. Components Lett.* **2017**, *27*, 476.
[211] G. Carchon, B. Nanwelaers, *IEEE Trans. Microw. Theory Tech.* **2000**, *48*, 316.
[212] T. Kodera, D. L. Sounas, C. Caloz, *Appl. Phys. Lett.* **2011**, *99*, 10.
[213] N. A. Estep, D. L. Sounas, A. Alu, *IEEE Trans. Microw. Theory Tech.* **2016**, 1.
[214] B.-I. Popa, S. A. Cummer, *Phys. Rev. B* **2012**, *85*, 205101.
[215] L. Chen, Q. Ma, H. B. Jing, H. Y. Cui, Y. Liu, T. J. Cui, *Phys. Rev. Appl.* **2019**, *11*, 054051.
[216] B. Peng, S. K. Özdemir, F. Lei, F. Monifi, M. Gianfreda, G. L. Long, S. Fan, F. Nori, C. M. Bender, L. Yang, *Nat. Phys.* **2014**, *10*, 394.
[217] A. Krasnok, A. Alù, *Proc. IEEE* **2019**, 1.
[218] K. G. Makris, R. El-Ganainy, D. N. Christodoulides, Z. H. Musslimani, *Phys. Rev. Lett.* **2008**, *100*, 103904.
[219] C. E. Rüter, K. G. Makris, R. El-Ganainy, D. N. Christodoulides, M. Segev, D. Kip, *Nat. Phys.* **2010**, *6*, 192.
[220] Y. Liu, T. Hao, W. Li, J. Capmany, N. Zhu, M. Li, *Light Sci. Appl.* **2018**, *7*, 38.
[221] S. Longhi, *Epl* **2017**, *120*, 64001.
[222] R. El-Ganainy, K. G. Makris, M. Khajavikhan, Z. H. Musslimani, S. Rotter, D. N. Christodoulides, *Nat. Phys.* **2018**, *14*, 11.
[223] S. V. Suchkov, A. A. Sukhorukov, J. Huang, S. V. Dmitriev, C. Lee, Y. S. Kivshar, *Laser Photonics Rev.* **2016**, *10*, 177.
[224] R. Thomas, H. Li, F. M. Ellis, T. Kottos, *Phys. Rev. A* **2016**, *94*, 1.
[225] J. Zhang, B. Peng, Ş. K. Özdemir, Y. Liu, H. Jing, X. Lü, Y. Liu, L. Yang, F. Nori, *Phys. Rev. B* **2015**, *92*, 115407.
[226] L. Del Bino, J. M. Silver, S. L. Stebbings, P. Del'Haye, *Sci. Rep.* **2017**, *7*, 43142.
[227] T. T. Koutserimpas, R. Fleury, *Phys. Rev. Lett.* **2018**, *120*, 87401.
[228] Y. Choi, C. Hahn, J. W. Yoon, S. H. Song, P. Berini, *Nat. Commun.* **2017**, *8*, 14154.
[229] H. Li, A. Mekawy, A. Krasnok, A. Alù, *Phys. Rev. Lett.* **2020**, *124*, 193901.
[230] S. Lepeshov, A. Krasnok, *Optica* **2020**, *7*, 1024.
[231] A. Kord, Magnetless Circulators Based on Linear Time-Varying Circuits, UT Austin, **2019**.
[232] A. Kord, D. L. Sounas, A. Alu, *IEEE Trans. Microw. Theory Tech.* **2018**, *66*, 2731.
[233] A. Kord, H. Krishnaswamy, A. Alù, *Phys. Rev. Appl.* **2019**, *12*, 024046.
[234] A. Kord, D. L. Sounas, Z. Xiao, A. Alu, *IEEE Trans. Microw. Theory Tech.* **2018**, *66*, 5472.
[235] A. Kord, M. Tymchenko, D. L. Sounas, H. Krishnaswamy, A. Alu, *IEEE Trans. Microw. Theory Tech.* **2019**, *67*, 2649.
[236] C. Cassella, G. Michetti, M. Pirro, Y. Yu, A. Kord, D. L. Sounas, A. Alu, M. Rinaldi, *IEEE Trans. Ultrason. Ferroelectr. Freq. Control* **2019**, *66*, 1814.
[237] D. Correas-Serrano, A. Alù, J. S. Gomez-Diaz, *Phys. Rev. B* **2018**, *98*, 165428.
[238] D. Correas-Serrano, J. S. Gomez-Diaz, D. L. Sounas, Y. Hadad, A. Alvarez-Melcon, A.

Alu, *IEEE Antennas Wirel. Propag. Lett.* **2016**, *15*, 1529.
[239] M.-A. Miri, E. Verhagen, A. Alù, *Phys. Rev. A* **2017**, *95*, 053822.
[240] M. A. Miri, F. Ruesink, E. Verhagen, A. Alù, *Phys. Rev. Appl.* **2017**, *7*, 1.
[241] E. Verhagen, A. Alù, *Nat. Phys.* **2017**, *13*, 922.
[242] X. Zhao, K. Wu, C. Chen, T. G. Bifano, S. W. Anderson, X. Zhang, *Adv. Sci.* **2020**, 2001443.
[243] S. Abdollahramezani, O. Hemmatyar, H. Taghinejad, A. Krasnok, Y. Kiarashinejad, M. Zandehshahvar, A. Alù, A. Adibi, *Nanophotonics* **2020**, *9*, 1189.
[244] S. Lepeshov, A. Krasnok, A. Alù, *ACS Photonics* **2019**, *6*, DOI 10.1021/acsphotonics.9b00674.
[245] C. Wan, E. H. Horak, J. King, J. Salman, Z. Zhang, Y. Zhou, P. Roney, B. S. Gundlach, S. Ramanathan, R. H. Goldsmith, M. A. Kats, P. Roney, B. S. Gundlach, S. Ramanathan, R. H. Goldsmith, M. A. Kats, *ACS Photonics* **2018**, *5*, 2688.
[246] D. A. Dobrykh, A. V. Yulin, A. P. Slobozhanyuk, A. N. Poddubny, Y. S. Kivshar, *Phys. Rev. Lett.* **2018**, *121*, 163901.
[247] D. Filonov, Y. Kramer, V. Kozlov, B. A. Malomed, P. Ginzburg, *Appl. Phys. Lett.* **2016**, *109*, 111904.
[248] D. L. Sounas, A. Alu, *IEEE Antennas Wirel. Propag. Lett.* **2018**, *17*, 1958.
[249] D. L. Sounas, A. Alù, *Phys. Rev. Lett.* **2017**, *118*, 154302.
[250] D. L. Sounas, J. Soric, A. Alù, *Nat. Electron.* **2018**, *1*, 113.
[251] B. Jin, C. Argyropoulos, *Phys. Rev. Appl.* **2020**, *13*, 054056.
[252] K. Y. Yang, J. Skarda, M. Cotrufo, A. Dutt, G. H. Ahn, M. Sawaby, D. Vercruysse, A. Arbabian, S. Fan, A. Alù, J. Vučković, *Nat. Photonics* **2020**, *14*, 369.
[253] A. Y. Piggott, E. Y. Ma, L. Su, G. H. Ahn, N. V. Sapra, D. Vercruysse, A. M. Netherton, A. S. P. Khope, J. E. Bowers, J. Vučković, *ACS Photonics* **2020**, *7*, 569.
[254] A. E. Krasnok, A. E. Miroshnichenko, P. A. Belov, Y. S. Kivshar, *JETP Lett.* **2011**, *94*, 593.
[255] A. E. Krasnok, A. E. Miroshnichenko, P. A. Belov, Y. S. Kivshar, *Opt. Express* **2012**, *20*, 20599.
[256] D. S. Filonov, A. E. Krasnok, A. P. Slobozhanyuk, P. V. Kapitanova, E. A. Nenasheva, Y. S. Kivshar, P. A. Belov, *Appl. Phys. Lett.* **2012**, *100*, 201113.
[257] A. E. Krasnok, I. S. Maksymov, A. I. Denisyuk, P. A. Belov, A. E. Miroshnichenko, C. R. Simovski, Y. S. Kivshar, *Physics-Uspekhi* **2013**, *56*, 539.
[258] A. Krasnok, S. Glybovski, M. Petrov, S. Makarov, R. Savelev, P. Belov, C. Simovski, Y. Kivshar, *Appl. Phys. Lett.* **2016**, *108*, 211105.
[259] A. E. Krasnok, D. S. Filonov, C. R. Simovski, Y. S. Kivshar, P. A. Belov, *Appl. Phys. Lett.* **2014**, *104*, DOI 10.1063/1.4869817.
[260] A. E. Krasnok, C. R. Simovski, P. A. Belov, Y. S. Kivshar, *Nanoscale* **2014**, *6*, 7354.
[261] P. A. Belov, A. E. Krasnok, D. S. Filonov, C. R. Simovski, Y. S. Kivshar, *IOP Conf. Ser. Mater. Sci. Eng.* **2014**, *67*, 012008.
[262] R. S. Savelev, D. S. Filonov, P. V. Kapitanova, A. E. Krasnok, A. E. Miroshnichenko, P. A. Belov, Y. S. Kivshar, *Appl. Phys. Lett.* **2014**, *105*, 181116.
[263] D. S. Filonov, A. P. Slobozhanyuk, A. E. Krasnok, P. A. Belov, E. A. Nenasheva, B. Hopkins, A. E. Miroshnichenko, Y. S. Kivshar, *Appl. Phys. Lett.* **2014**, *104*, DOI 10.1063/1.4858969.
[264] A. Krasnok, S. Li, S. Lepeshov, R. Savelev, D. G. Baranov, A. Alú, *Phys. Rev. Appl.*

**2018**, *9*, 14015.
[265] A. Krasnok, M. Tymchenko, A. Alù, *Mater. Today* **2018**, *21*, 8.
[266] R. S. Savelev, O. N. Sergaeva, D. G. Baranov, A. E. Krasnok, A. Alù, *Phys. Rev. B* **2017**, *95*, 235409.
[267] S. Fan, W. Suh, J. D. Joannopoulos, *J. Opt. Soc. Am. A* **2003**, *20*, 569.
[268] H. J. Kimble, *Nature* **2008**, *453*, 1023.
[269] D. Roy, *Sci. Rep.* **2013**, *3*, 2337.
[270] F. Fratini, E. Mascarenhas, L. Safari, J.-P. Poizat, D. Valente, A. Auffèves, D. Gerace, M. F. Santos, *Phys. Rev. Lett.* **2014**, *113*, 243601.
[271] J. Dai, A. Roulet, H. N. Le, V. Scarani, *Phys. Rev. A* **2015**, *92*, 063848.
[272] F. Fratini, R. Ghobadi, *Phys. Rev. A* **2016**, *93*, 023818.
[273] C. Müller, J. Combes, A. R. Hamann, A. Fedorov, T. M. Stace, *Phys. Rev. A* **2017**, *96*, 053817.
[274] M. Scheucher, A. Hilico, E. Will, J. Volz, A. Rauschenbeutel, *Science (80-. ).* **2016**, *354*, 1577.
[275] A. Javadi, D. Ding, M. H. Appel, S. Mahmoodian, M. C. Löbl, I. Söllner, R. Schott, C. Papon, T. Pregnolato, S. Stobbe, L. Midolo, T. Schröder, A. D. Wieck, A. Ludwig, R. J. Warburton, P. Lodahl, *Nat. Nanotechnol.* **2018**, *13*, 398.
[276] Y. Shen, M. Bradford, J.-T. Shen, *Phys. Rev. Lett.* **2011**, *107*, 173902.
[277] P. Lodahl, S. Mahmoodian, S. Stobbe, *Rev. Mod. Phys.* **2015**, *87*, 347.
[278] A. Laucht, S. Pütz, T. Günthner, N. Hauke, R. Saive, S. Frédérick, M. Bichler, M.-C. Amann, A. W. Holleitner, M. Kaniber, J. J. Finley, *Phys. Rev. X* **2012**, *2*, 011014.
[279] A. Goban, C.-L. Hung, S.-P. Yu, J. D. Hood, J. A. Muniz, J. H. Lee, M. J. Martin, A. C. McClung, K. S. Choi, D. E. Chang, O. Painter, H. J. Kimble, *Nat. Commun.* **2014**, *5*, 3808.
[280] I. Söllner, S. Mahmoodian, S. L. Hansen, L. Midolo, A. Javadi, G. Kiršanske, T. Pregnolato, H. El-Ella, E. H. Lee, J. D. Song, S. Stobbe, P. Lodahl, *Nat. Nanotechnol.* **2015**, *10*, 775.
[281] E. J. Lenferink, G. Wei, N. P. Stern, *Opt. Express* **2014**, *22*, 16099.
[282] K. Xia, G. Lu, G. Lin, Y. Cheng, Y. Niu, S. Gong, J. Twamley, *Phys. Rev. A* **2014**, *90*, 043802.
[283] P. A. Dmitriev, D. G. Baranov, V. A. Milichko, S. V. Makarov, I. S. Mukhin, A. K. Samusev, A. E. Krasnok, P. A. Belov, Y. S. Kivshar, *Nanoscale* **2016**, *8*, 9721.
[284] K. Frizyuk, M. Hasan, A. Krasnok, A. Alú, M. Petrov, *Phys. Rev. B* **2018**, *97*, 085414.
[285] A. Krasnok, S. Lepeshov, A. Alú, *Opt. Express* **2018**, *26*, 15972.
[286] H. Zeng, J. Dai, W. Yao, D. Xiao, X. Cui, *Nat. Nanotechnol.* **2012**, *7*, 490.
[287] K. F. Mak, K. He, J. Shan, T. F. Heinz, *Nat. Nanotechnol.* **2012**, *7*, 494.
[288] T. Cao, G. Wang, W. Han, H. Ye, C. Zhu, J. Shi, Q. Niu, P. Tan, E. Wang, B. Liu, J. Feng, *Nat. Commun.* **2012**, *3*, 885.
[289] A. Krasnok, *Nat. Nanotechnol.* **2020**, *15*, 893.